\documentclass[fleqn,usenatbib]{mnras}
\usepackage{aas_macros}
\usepackage{abbreviations}
\usepackage{amssymb}
\usepackage{bigdelim}
\usepackage{bm}
\usepackage{cprotect}
\usepackage{etoolbox}
\usepackage{gensymb}
\usepackage{graphicx}
\usepackage{hyperref}
\usepackage{mathtools}
\usepackage{multirow}
\usepackage{multicol}
\usepackage{subcaption}
\usepackage{threeparttable}
\usepackage{upgreek}
\usepackage{xcolor}
\usepackage[figuresright]{rotating}
\usepackage{newtxtext,newtxmath}
\usepackage[T1]{fontenc}

\definecolor{applegreen}{rgb}{0.55, 0.71, 0.0}
\def\xmm{\textit{XMM-Newton}}

\hypersetup{
  colorlinks   = true, 
  urlcolor     = blue, 
  linkcolor    = red, 
  citecolor   = blue 
}

\title[A MeerKAT-meets-LOFAR study of MS~1455.0$+$2232]{A MeerKAT-meets-LOFAR Study of MS~1455.0$+$2232: A 590~kiloparsec `Mini'-Halo in a Sloshing Cool-Core Cluster}

\author[C.~J.~Riseley et al.]{C.~J.~Riseley$^{1,2,3}$\thanks{Corresponding author email: \url{christopher.riseley@unibo.it}}, \hspace{0.001cm} K.~Rajpurohit$^{1,2,4}$, F.~Loi$^5$, A.~Botteon$^6$, R.~Timmerman$^6$, 
N.~Biava$^{1,2}$, A.~Bonafede$^{1,2}$,
\newauthor
E.~Bonnassieux$^{1,2}$, G.~Brunetti$^2$, T.~En{\ss}lin$^7$, G.~Di~Gennaro$^8$, A.~Ignesti$^9$, T.~Shimwell$^6$, C.~Stuardi$^{1,2}$,
\newauthor
T.~Vernstrom$^{3,10}$, R.~J.~van~Weeren$^6$
\\
$^1$ Dipartimento di Fisica e Astronomia, Universit\`a degli Studi di Bologna, via P. Gobetti 93/2, 40129 Bologna, Italy \\ 
$^2$ INAF -- Istituto di Radioastronomia, via P. Gobetti 101, 40129 Bologna, Italy \\ 
$^3$ CSIRO Space \& Astronomy, PO Box 1130, Bentley, WA 6102, Australia \\ 
$^4$ Th\"{u}ringer Landessternwarte, Sternwarte 5, 07778 Tautenburg, Germany \\ 
$^5$ INAF-Osservatorio Astronomico di Cagliari, Via della Scienza 5, 09047 Selargius, CA, Italy \\ 
$^6$ Leiden Observatory, Leiden University, P.O. Box 9513, 2300 RA Leiden, The Netherlands \\ 
$^7$ Max Planck Institute for Astrophysics, Karl-Schwarzschild-Stra{\ss}e 1, 85748 Garching, Germany \\ 
$^8$ Hamburger Sternwarte, Universität Hamburg, Gojenbergsweg 112, 21029 Hamburg, Germany \\ 
$^9$ INAF -- Astronomical Observatory of Padova, vicolo dell'Osservatorio 5, IT-35122 Padova, Italy \\ 
$^{10}$ ICRAR, The University of Western Australia, 35 Stirling Hw, 6009 Crawley, Australia 
}

\date{Accepted 2022 March 08. Received 2022 March 07; in original form 2021 December 13.}

\pubyear{2022}

\begin{document}
\label{firstpage}
\pagerange{\pageref{firstpage}--\pageref{lastpage}}
\maketitle

\begin{abstract}
Radio mini-haloes are poorly-understood, moderately-extended diffuse radio sources that trace the presence of magnetic fields and relativistic electrons on scales of hundreds of kiloparsecs, predominantly in relaxed clusters. With relatively few confirmed detections to-date, many questions remain unanswered. This paper presents new radio observations of the galaxy cluster MS~1455.0$+$2232 performed with MeerKAT (covering the frequency range 872$-$1712~MHz) and LOFAR (covering 120$-$168~MHz), the first results from a homogeneously selected mini-halo census. We find that this mini-halo extends for $\sim590$~kpc at 1283~MHz, significantly larger than previously believed, and has a flatter spectral index ($\alpha = -0.97 \pm 0.05$) than typically expected. Our X-ray analysis clearly reveals a large-scale ($254~$kpc) sloshing spiral in the intracluster medium. We perform a point-to-point analysis, finding a tight single correlation between radio and X-ray surface brightness with a super-linear slope of $b_{\rm 1283~MHz} = 1.16^{+0.06}_{-0.07}$ and $b_{\rm 145~MHz} = 1.15^{+0.09}_{-0.08}$; this indicates a strong link between the thermal and non-thermal components of the intracluster medium. Conversely, in the spectral index/X-ray surface brightness plane, we find that regions inside and outside the sloshing spiral follow different correlations. We find compelling evidence for multiple sub-components in this mini-halo for the first time. While both the turbulent (re-)acceleration and hadronic scenarios are able to explain some observed properties of the mini-halo in MS~1455.0$+$2232, neither scenario is able to account for all the evidence presented by our analysis.
\end{abstract}

\begin{keywords}
Galaxies: clusters: general -- galaxies: clusters: individual: MS~1455.0$+$2232 -- radio continuum: general -- X-rays: galaxies: clusters -- galaxies: clusters: intracluster medium
\end{keywords}

\section{Introduction}
The intracluster medium (ICM) constitutes some 80 per cent of the baryonic mass of a galaxy cluster, comprising a hot ($T\sim10^7-10^8$~K) tenuous ($n_e\sim10^{-3}$~cm$^{-3}$) plasma, which emits at X-ray wavelengths via Bremsstrahlung emission. During cluster merger events, tremendous quantities of energy \citep[$\sim10^{64}$~erg;][]{Ferrari2008} are deposited into the ICM via shocks and turbulence. Many merging clusters are known to host embedded radio galaxies which, when resolved, show clear evidence of interaction with the ICM \citep[e.g.][]{Mao2009,Douglass2011,Pratley2013,Owen2014,vanWeeren2017,Wilber2018,Clarke2019,Botteon2020b,Botteon2021,Bonafede2021,deGasperin2021}. On larger scales, many merging clusters host spectacular sources of diffuse radio emission, broadly classified as `radio haloes' and `radio relics' \citep[see][for a recent review]{vanWeeren2019}. There is a third class of diffuse radio source -- `radio mini-haloes', a term first introduced by \cite{Feretti1996} -- which have to-date been detected primarily in relaxed clusters, suggesting that major mergers are not the key ingredient in generating a mini-halo. Given the typically short synchrotron lifetime of relativistic electrons ($\tau\lesssim10^8$~yr) compared to the spatial scales over which the diffuse radio emission is detected, all three broad classes of non-thermal phenomena associated with the ICM require in-situ acceleration \citep{Jaffe1977}. 

Mini-haloes remain the most poorly-understood class as they are relatively rare objects, with only some 32 currently known, compared to around $\sim72$ clusters hosting radio relics and $\sim95$ hosting radio haloes \citep{vanWeeren2019}. We note however that with the advent of many next-generation radio interferometers, the known population of diffuse radio sources in clusters of galaxies is expanding dramatically \citep[for example,][]{Govoni2019,Botteon2020a,Hodgson2021,Osinga2021,vanWeeren2021,Duchesne2021c,Duchesne2022,Knowles2022}. Mini-haloes are moderately-extended (typically $\sim0.1-0.3$~Mpc), steep-spectrum ($\alpha \lesssim -1$, where $\alpha$ is the spectral index, relating flux density $S$ to observing frequency $\nu$ as $S \propto \nu^{\alpha})$ objects that are centrally-located in relaxed clusters which also host a radio source associated with the nucleus of the brightest cluster galaxy (BCG).

Much discussion has occurred over the past two decades in the literature regarding the nature of the acceleration mechanism responsible for powering the relativistic electrons that generate mini-haloes. Two principal mechanisms exist: turbulent (re-)acceleration \citep[][]{Gitti2002,ZuHone2013} and hadronic processes \citep[][]{Pfrommer2004}.

The turbulent (re-)acceleration scenario implies a careful balance between injecting sufficient turbulence to induce large-scale gas sloshing motions, while also leaving the relaxed cool core intact. Simulations generally suggest that minor and/or off-axis mergers can provide the necessary energy input \citep[e.g.][]{ZuHone2013}, and observations have shown that mini-haloes tend to be bounded by cold fronts \citep[see for example][]{Mazzotta2001,Mazzotta2001b,Mazzotta2008,Ghizzardi2010,Rossetti2013}, suggesting sloshing motions that would introduce turbulence. Indeed, under reasonable conditions, theoretical predictions are able to match the observed variations in spectral index \citep[][]{Giacintucci2014b}. Finally, the co-location of many mini-haloes with the radio BCG may suggest a natural source for the seed population of mildly-relativistic electrons that would be re-accelerated to the relativistic regime \citep[e.g.][]{Fujita2007}. The correlations between mini-halo and BCG radio power, as well as mini-halo power and X-ray cavity power, provide further observational support for this scenario \citep[][]{Bravi2016,RichardLaferriere2020}.

In the hadronic scenario, cosmic-ray electrons (CRe) are continuously injected via inelastic collisions between cosmic-ray protons (CRp) and the cluster thermal proton population. CRp are expected to persist at some level throughout the cluster volume, being injected by active galactic nuclei (AGN) and supernovae processes \citep[for a comprehensive review, see][]{Brunetti2014}. For giant radio haloes, purely-hadronic models cannot work: one of the arguments against this scenario comes from \emph{Fermi} upper limits to $\gamma$-ray emission from giant halo clusters, which are too constraining \citep[e.g.][]{Brunetti2017}. However, it has been recently shown by \cite{Ignesti2020} that the current limits from \emph{Fermi} are compatible with the hadronic scenario in mini-halo clusters.

Hadronic models generally predict a smaller range of observed spectral index values \citep[$-1.2 \lesssim \alpha \lesssim -1$;][]{Pfrommer2004,Brunetti2014,ZuHone2015}. Conversely, the turbulent (re-)acceleration scenario can yield a much broader range, including ultra-steep spectral indices \citep[$\alpha\lesssim-1.5$;][]{Brunetti2008,ZuHone2013,Brunetti2014}. Under the assumption of homogeneous conditions throughout the mini-halo, the turbulent (re-)acceleration scenario should lead to an observable spectral break toward high frequencies (due to the cutoff in the electron energy distribution). The presence of a high-frequency spectral break remains an open question, however. Few studies have the necessary high-quality high-frequency data, although some mini-haloes appear to show power-law spectra up to at least 10$-$20~GHz \citep{Timmerman2021,Perrott2021}.

Despite the differences in formation scenario, both scenarios do share some commonalities: the important role of cluster-member AGN (particularly the BCG) in providing seed particles, and the likely connection between the thermal and non-thermal components in the ICM. As such, correlations are expected between various probes of these components (e.g. radio and X-ray surface brightness). Recent work by \cite{Ignesti2020} attempted to investigate the statistical properties of the radio/X-ray connection for seven mini-haloes. Contrary to what is typically found in point-to-point correlation studies of giant haloes, \citeauthor{Ignesti2020} found that mini-haloes tend to exhibit a super-linear correlation between radio surface brightness and X-ray surface brightness, i.e. $b>1$ for $I_{\rm{radio}} \propto I_{\rm{X-ray}}^{b}$. However, their sample also showed significant scatter and relatively large uncertainty in the correlation slope $b$, largely due to the resolution of the available data \citep[see also discussion by][]{Bruno2021}. Additionally, no statistically-significant sample of high-quality mini-halo spectra exist, meaning no robust conclusions can be drawn regarding the spectral properties or the underlying acceleration mechanism.

Recently, the picture has become far more complex with new highly-sensitive long-wavelength observations from the LOw-Frequency ARray \citep[LOFAR;][]{vanHaarlem2013} revealing diffuse radio emission far outside the cool core of a handful of mini-halo clusters \citep[][]{Savini2018,Savini2019,Biava2021}. These complex (ultra-)steep spectrum sources challenge the strict divide that has historically existed between mini-haloes and larger-scale haloes, and hint at different particle acceleration mechanisms being dominant on different scales. We also note that mini-haloes are not the only objects which show evidence of such complexity. \cite{Bonafede2014} detected a giant ($\sim1.1$~Mpc) radio halo in the cool-core cluster CL~1821$+$643. Additionally, the `radio halo' in Abell~2142 exhibits evidence of substructure, including a brighter central component of mini-halo-like emission that blends into larger-scale diffuse halo-like emission \citep{Venturi2017}.

We are using deep observations with MeerKAT \citep{Jonas2016}\footnote{The technical capabilities of MeerKAT are also discussed further by \cite{Camilo2018} and \cite{Mauch2020}} covering the frequency range 872$-$1712~MHz, and the LOFAR High Band Antenna (HBA), covering the frequency range 120$-$168~MHz, to perform a multi-frequency census of all known radio mini-haloes in the Declination range $-1\degree$ to $+30\degree$. The MeerKAT project code is MKT-20126 (P.I. Riseley). This multi-frequency data is augmented by archival X-ray data from \textit{Chandra} and \xmm{}. The primary goal of our census is to determine the nature of particle acceleration mechanisms at play in mini-haloes using the first uniformly-constructed mini-halo sample. This paper presents new results on the mini-halo in the cluster MS~1455.0$+$2232, carried out as a pilot study for our census. 

In the next sub-section, we will briefly summarise what is currently known about MS~1455.0$+$2232. The remainder of this paper is divided as follows: we discuss the observations and data reduction in \S\ref{sec:observations}, we present our results in \S\ref{sec:results} and analyse them in \S\ref{sec:analysis}. In \S\ref{sec:discussion} we discuss our results in the context of the two principal acceleration scenarios. We draw our conclusions in \S\ref{sec:conclusions}. Throughout, we assume a $\Lambda$CDM cosmology of H$_0 = 73 ~ \rm{km} ~ \rm{s}^{-1} ~ \rm{Mpc}^{-1}$, $\Omega_{\rm{m}} = 0.27$, $\Omega_{\rm{vac}} = 0.73$. Given our cosmology, at the redshift of MS~1455.0$+$2232 \citep[$z=0.258$;][]{Allen1992}, an angular distance of 1~arcsec corresponds to a physical distance of 3.859~kpc. All uncertainties are quoted at the $1\upsigma$ level.

\begin{figure}
\begin{center}
\includegraphics[width=\linewidth]{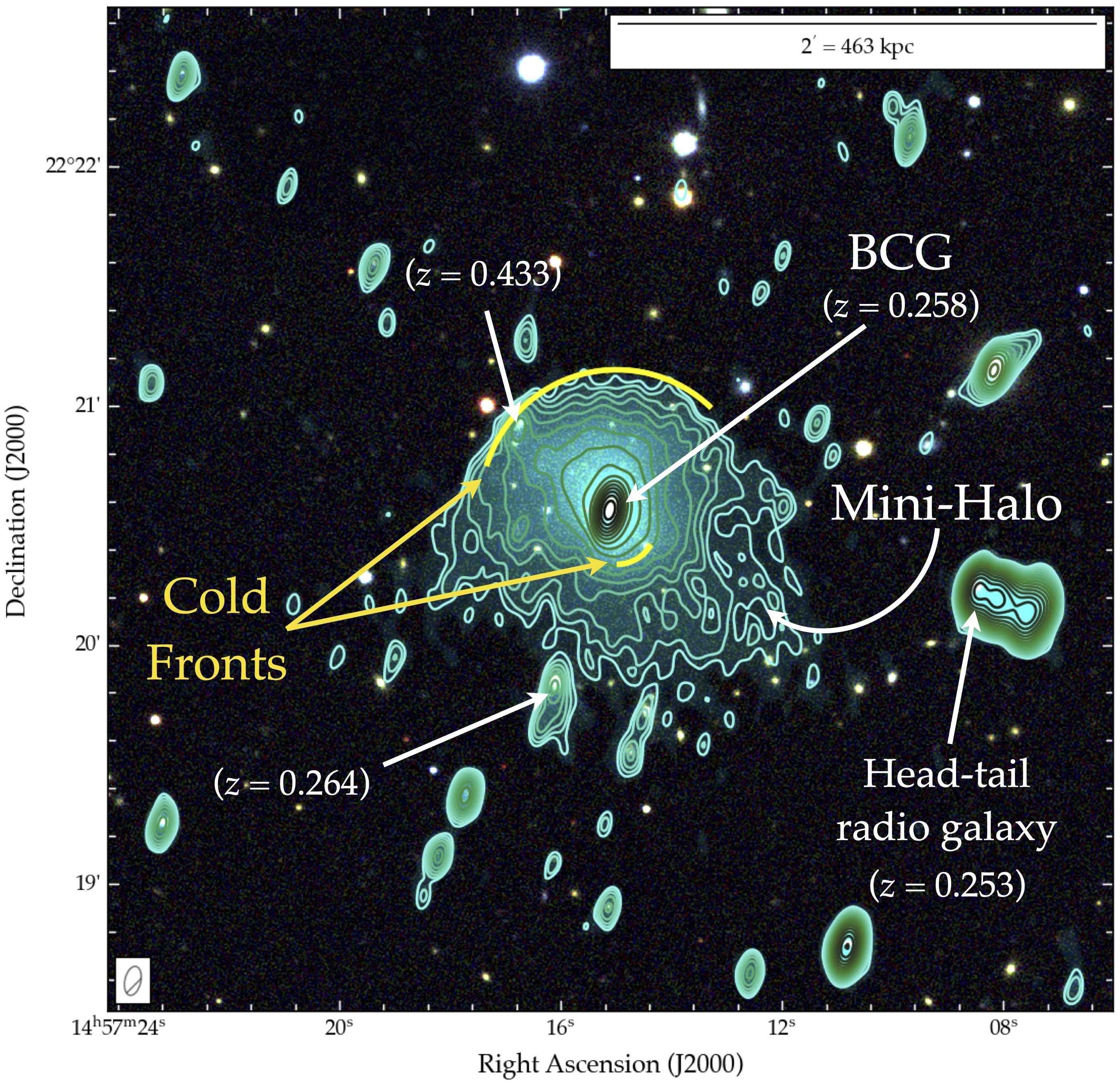}
\cprotect\caption{Colour-composite image of MS~1455.0$+$2232, showing 1283~MHz radio surface brightness measured by MeerKAT at $7.4\times4.0$~arcsec resolution (teal color and contours) overlaid on an optical RGB image constructed using $i$-, $r$- and $g$-band images from SDSS DR16 \citep{Ahumada2020}. Contours start at $4\sigma$ and scale by a factor $\sqrt{2}$, where $\sigma = 5.9~\upmu$Jy beam$^{-1}$; see Table~\ref{tab:img_summary}. The two known cold fronts identified by \cite{Mazzotta2001} are shown with yellow arcs; the redshifts of selected galaxies are also shown.}
\label{fig:Z7160_rgb}
\end{center}
\end{figure}

\subsection{The Mini-Halo in MS~1455.0+2232}
MS~1455.0$+$2232 (also commonly referred to in the literature as Z7160 or ZwCl~1454.8$+$2233) is a cool-core cluster hosting a pair of cold fronts on opposing sides of the cluster centre \citep[][]{Mazzotta2001,Mazzotta2008}. The position of these cold fronts relative to the larger-scale emission from MS~1455.0$+$2232 is presented in Figure~\ref{fig:Z7160_rgb}. We show an illustrative colour-composite image, with our 1283~MHz MeerKAT map overlaid on an optical RGB image reconstructed using $i$-, $r$- and $g$-band images from the Sloan Digital Sky Survey (SDSS) Data Release 16 \citep{Ahumada2020}.

This cluster has a mass of $M_{\rm{500}} = (3.5 \pm 0.4) \times 10^{14}~{\rm{M}}_{\odot}$, estimated by \cite{Giacintucci2017} using the $T_{\rm{X}}/M_{500}$ relation of \cite{Vikhlinin2009}. This is relatively low mass for known mini-halo clusters -- for example, it is the second-least-massive cluster in the sample of \cite{Giacintucci2019}. We do however note some discrepancy in the literature regarding the cluster mass: \cite{Rossetti2017} derive a much greater value of $M_{\rm{500}} = 7 \times 10^{14}~{\rm{M}}_{\odot}$ using the $L_{\rm X}/M_{\rm 500}$ of \cite{Pratt2009}. The central entropy for MS~1455.0$+$2232 is 16.88~keV~cm$^{2}$ \citep{Cavagnolo2009}.

The mini-halo in MS~1455.0$+$2232 was first reported by \cite{Venturi2008} using 610~MHz observations performed with the Giant Metrewave Radio Telescope \citep[GMRT;][]{swarup1990}. \citeauthor{Venturi2008} identified it as a `core-halo' structure, measuring a largest angular size of around 90~arcsec (347~kpc with our cosmology). More recently, \cite{Giacintucci2019} present a re-analysis of archival narrow-band 1.4~GHz Very Large Array (VLA) A- and C-configuration observations. The mini-halo is only partially resolved in their C-configuration image, and appears less extended ($\sim60$~arcsec in diameter, or 231~kpc with our cosmology) than at lower frequencies. \citeauthor{Giacintucci2019} estimate a spectral index of $\alpha_{610~\rm{MHz}}^{1.4~\rm{GHz}} = -1.45 \pm 0.22$.

\section{Observations \& Data Reduction}\label{sec:observations}

\subsection{Radio: MeerKAT}
MS~1455.0$+$2232 was observed with the MeerKAT telescope on 2021~April~25 (SB~20210301-0026). Observations were performed using the L-band receivers, with 4096 channels covering a frequency range 872$-$1712~MHz. Following standard MeerKAT practice, bandpass calibrators were observed roughly every three hours during the observing run. These were the bright radio sources PKS~B0407$-$658 and PKS~B1934$-$638, the latter of which was also used to set the absolute flux density scale. The compact radio source J1445$+$0958 was observed for two minutes at a half-hour cadence to provide an initial calibration of the time-varying instrumental gains. The total integration time on MS~1455.0$+$2232 was 5.5~hours.

To calibrate the electric vector position angle (EVPA), we included two five-minute scans on the known polarisation calibrator 3C~286, separated by a broad parallactic angle range. While mini-haloes are expected to be largely depolarised, no statistical study of their polarisation properties has been performed to-date. We will examine the polarisation properties of our full sample of mini-haloes (as well as radio sources of interest in the wider field) in future works.

Initial calibration and flagging was performed using the Containerized Automated Radio Astronomy Calibration (\texttt{CARACal}) pipeline\footnote{\url{https://github.com/caracal-pipeline/caracal}} \citep{Jozsa2020,Jozsa2021}, which uses the Stimela framework \citep{Makhathini2018} as a wrapper around standard Common Astronomy Software Application (\textsc{CASA}) calibration tasks. We used \texttt{CARACal} to perform full-polarisation calibration, although this paper details exclusively radio continuum results. 

The \texttt{CARACal} pipeline also incorporates various flaggers. During pipeline execution, we flagged shadowed antennas as well as specific channel ranges corresponding to the MeerKAT bandpass edges and bands that are known to be dominated by radio frequency interference (RFI). We also applied automatic sum-threshold flagging to our calibrator data using the \verb|tfcrop| algorithm. After a first pass through flagging and initial calibration, we re-flagged our data using \verb|tfcrop| before re-deriving all calibration tables---delays, bandpass, parallel-hand and cross-hand gains, and instrumental leakage---to refine our solutions, before applying these calibration tables to our target. 

An initial round of flagging was performed on our target visibilities using the \texttt{Tricolour} flagger\footnote{\url{https://github.com/ska-sa/tricolour}} (Hugo et al., in press 2021) to execute a first (relatively shallow) pass of automated sum-threshold flagging within \texttt{CARACal}. We then generated an initial sky model using \texttt{DDFacet} \citep{Tasse2018}, which was subtracted from our data before running a second pass of \texttt{Tricolour} flagging on the residual data. Finally, we averaged our data by a factor of eight in frequency, yielding 512 output channels of 1.67~MHz width, before proceeding to self-calibration. 

All self-calibration was carried out using the \texttt{killMS} software package \citep[\texttt{kMS};][]{Tasse2014,Smirnov2015}. We performed three rounds of direction-independent (DI) phase-only self-calibration and two rounds of DI amplitude-and-phase self-calibration. Imaging was again performed with \texttt{DDFacet} using \texttt{robust} $=-0.5$ weighting \citep{Briggs1995} in order to achieve a balance between resolution, signal-to-noise, and sensitivity to diffuse radio emission. We used the sub-space deconvolution \citep[\texttt{SSD};][]{Tasse2018} algorithm in order to more accurately reconstruct the emission from the various resolved radio sources across the field of view. After five rounds of DI self-calibration, our solutions had largely converged, and we inspected our image for residual direction-dependent (DD) errors.

Throughout the self-calibration process, we used a quality-based weighting scheme \citep[see][for details]{Bonnassieux2018} to weight our calibration solutions. This scheme makes use of gain statistics to estimate the relative quality of each solution, and weight each corrected visibility accordingly. In general, this results in faster self-calibration convergence. These weights are computed by default by \texttt{kMS} after each solve, as minimal computational overhead is required.

While we had achieved a generally high dynamic range after DI self-calibration, we noted some residual DD errors in the wider field. To correct for these, we tessellated the sky into 16 directions and performed two rounds of DD-calibration, applying gains in both amplitude and phase during each imaging step. Once no further improvement was achieved from our self-calibration process, we generated an `extracted' dataset covering the immediate region around MS~1455.0$+$2232 by subtracting our best sky model of all sources more than 0.2~degree from the cluster. MeerKAT has a large field of view at L-band: the primary beam full-width at half-maximum (FWHM) is $\sim67$~arcmin at 1.28~GHz; \citep[][]{Mauch2020}. As our target occupies a small fraction of this field of view, the extraction step allowed us to re-image our target with different weighting schemes and \emph{uv}-selection ranges with manageable computational overhead.

\subsection{Radio: LOFAR}
MS~1455.0$+$2232 was observed with LOFAR as part of the LOFAR Two-Metre Sky Survey \citep[LoTSS;][]{Shimwell2017}. The nearest field centre (P225+22; 0.81~degree separation) was observed using the full International LOFAR Telescope \citep[ILT;][]{vanHaarlem2013} in \texttt{HBA\_DUAL\_INNER} mode for a total of 8~hours on 2020~May~03. These data cover the frequency range 120$-$169~MHz.

\subsubsection{Dutch LOFAR Array}
The data from the Dutch LOFAR array (Core and Remote stations, encompassing baselines out to $\sim80$~km) were processed using the LoTSS pipeline\footnote{\url{github.com/mhardcastle/ddf-pipeline/}}. The pipeline steps are discussed in depth by \cite{Shimwell2017,Shimwell2019} and \cite{Tasse2021}; however, in brief, this pipeline performs flagging, initial calibration and both DI and DD self-calibration using \texttt{killMS} and \texttt{DDFacet}. Following pipeline processing, an `extracted' dataset was created \citep[for details of the extraction process, see][]{vanWeeren2021} containing the region around MS~1455.0$+$2232.

\subsubsection{International LOFAR Telescope}
LoTSS observations are performed with the full ILT as standard, providing access to sub-arcsecond resolution images at 145~MHz. To perform the calibration using the complete LOFAR array, we first used the \textsc{Prefactor} software package \citep{vanWeeren2016, Williams2016, deGasperin2019} to calibrate the Dutch part of the array. Following initial RFI flagging, a model of the calibrator source (3C\,196) was used to correct for the polarization alignment, Faraday rotation, bandpass and finally clock offsets. With these calibration solutions applied to the data, a single phase-only cycle of self-calibration was performed on the Dutch stations using a sky model obtained from the first Alternative Data Release of the TIFR-GMRT Sky Survey \citep[TGSS-ADR1;][]{Intema2017}. Afterwards, the long-baseline pipeline discussed by \citet{Morabito2022} was employed to calibrate the international stations for ionospheric dispersive delays using a bright and compact source from the Long Baseline Calibrator Survey \citep[LBCS;][]{Jackson2016,Jackson2022}. These calibration solutions were transfered to the target source and applied to the data. To solve for the residual dispersive delay errors due to direction dependence of this effect, a final long-interval self-calibration routine is performed on a nearby source instead of the target source, due to the faintness of the target.

\subsection{Radio: Ancillary Data}
MS~1455.0$+$2232 has previously been observed with the GMRT using the legacy GMRT Software Backend (GSB) at 610~MHz, with 32~MHz bandwidth. Observations were carried out on 2005~October~03, for a total of 166~minutes on-source time. These observations were first published by \cite{Venturi2008} and have subsequently been used by other authors \citep[][]{Mazzotta2008,Ignesti2020}. We re-processed the raw visibilities from these observations using the Source Peeling and Atmospheric Modelling \citep[\texttt{SPAM};][]{Intema2009} pipeline in a fully automated manner, as described by \cite{Intema2017}.

We also sourced a 3~GHz mosaic image of MS~1455.0$+$2232 from the first two epochs of the Karl G. Jansky VLA Sky Survey \citep[VLASS;][]{Lacy2020} via the Canadian Initiative for Radio Astronomy Data Analysis (CIRADA) image cutout server. Finally, we searched the VLA data archive for historic VLA datasets covering MS~1455.0$+$2232. We retrieved observations at 1.4~GHz, 4.85~GHz and 8.46~GHz from projects AE117 (performed 1998~April~12), AE125 (performed 1999~January~16) and AE317 (performed 1994~March~12). These datasets were reduced using standard techniques in \textsc{CASA} v6.1.2.7. All ancillary datasets used in this work provide relatively limited sensitivity to the diffuse emission of the mini-halo, and thus their use was restricted to providing flux density measurements for the radio counterpart to the BCG. We do not present images from these archival datasets in this manuscript.

\subsection{Radio: Postprocessing}\label{sec:postprocessing}
After calibration, we used \texttt{WSclean} \citep{Offringa2014,Offringa2017} version 2.10.0\footnote{WSclean is available at \url{https://gitlab.com/aroffringa/wsclean}} to generate science-quality images from our extracted MeerKAT and LOFAR datasets, to ensure that no discrepancies were introduced by the use of different algorithms. 

When imaging with \texttt{WSclean}, we employed multi-scale clean in order to optimally recover diffuse radio emission. The \texttt{-auto-mask} and \texttt{-auto-threshold} functionality was used to automate the deconvolution process; with the well-filled \emph{uv}-plane of both our deep LOFAR and MeerKAT observations, each dataset has a well-behaved point spread function (PSF) with low-level sidelobes, and no significant spurious components were picked up by the auto-masked cleaning.

Given the large fractional bandwidth of both LOFAR and MeerKAT, we cleaned using the \texttt{-join-channels} and \texttt{-channels-out} options to improve the wide-band deconvolution. Stokes $I$ images were produced in 6 (24) sub-bands across the LOFAR HBA (MeerKAT) bandwidth, in addition to multi-frequency-synthesis (MFS) images at frequencies of 145~MHz for LOFAR and 1283~MHz for MeerKAT. We also employed an inner \emph{uv}-cut of $80\uplambda$ to exclude baselines between the two substations of each LOFAR HBA Core Station; to ensure a consistent maximum recoverable angular scale, we adopted the same inner \emph{uv}-cut for our MeerKAT imaging.

We then performed a further single round of DI self-calibration on our MeerKAT data using the LOFAR Default Pre-Processing Pipeline \citep[\texttt{DPPP};][]{vanDiepen2018}, solving for a full-Jones gain matrix. Further self-calibration did not yield appreciable improvement; similarly, further self-calibration of our extracted LOFAR dataset yielded no improvement in image quality. As such, we concluded that our calibration had converged and proceeded to create our final images.

Representative low- and high-resolution images were produced by varying the \texttt{robust} parameter between different \texttt{WSclean} imaging runs. We used \texttt{robust}~$=-0.5$ for low-resolution imaging and \texttt{robust}~$=-2.0$ (corresponding to \texttt{uniform} weighting) for high-resolution imaging.

\subsubsection{Flux Scaling}
The MeerKAT data presented in this work had the flux density scale set using observations of PKS~B1934$-$638, tied to the \cite{Reynolds1994} scale, which is itself tied to the \cite{Baars1977} scale. Observations with LOFAR are tied to the \cite{ScaifeHeald2012} flux scale. The flux scale of all ancillary VLA datasets used in this work was set using the \citeauthor{Baars1977} scale. The archival GMRT data were processed using the \texttt{SPAM} pipeline, which uses the \cite{ScaifeHeald2012} scale, so no conversion is required. When defining their flux scale, \citeauthor{ScaifeHeald2012} brought all measurements for their selected calibrator sources onto the \cite{RCB1973} scale, which is consistent with the \cite{Kellermann1966} scale above 325~MHz.

Table~7 of \cite{Baars1977} presents a set of factors for the ratio of their scale to other flux scales, including the \cite{Kellermann1966} scale. Thus, we used a polynomial fit to these ratios (performed in log-linear space) to derive the scaling factor required to bring our MeerKAT images into consistency with the LOFAR flux scale. The required conversion factor at the reference frequency of our MeerKAT observations ($\nu_{\rm{ref}} = 1283$~MHz) yielded by our polynomial fit was 0.968. We note that for all frequencies considered in this work, our $n$th-order polynomial fits converged to broadly consistent values for $4 \leq n \leq 10$; any error induced by imprecise flux scale conversion is thus small compared to the overall flux scale and measurement uncertainties.

To verify the flux scale of our LOFAR dataset, we used \texttt{WSclean} to generate an image at 6~arcsec resolution. This image was then compared with the well-verified flux density scale of LoTSS. This step involves extracting a catalogue from the target field, using the Python Blob Detection and Source Finder software \citep[\texttt{PyBDSF};][]{MohanRafferty2015}. The catalogue is then compared with the catalogue derived from the full-field LoTSS image, and filtered to select only compact sources with a signal-to-noise ratio at least seven. Finally, this routine performs a linear regression best-fit in the flux:flux plane in order to derive the bootstrap factor. The application of this technique is discussed in detail by \cite{Hardcastle2016} and \cite{Shimwell2019}. For our extracted LOFAR dataset, we require a bootstrap factor of 1.0899 to align with the LoTSS flux scale; for the full ILT image, we require a bootstrap factor 1.149. We adopt a typical 5 per cent uncertainty in the flux scale of our MeerKAT images; following \cite{Shimwell2022}, we also adopt a representative 10 per cent uncertainty for the flux scale of our LOFAR images.

\subsubsection{Source Subtraction and Final Imaging}
Several compact or partially-resolved sources are visible in the region of the mini-halo in MS~1455.0$+$2232, including the compact radio source associated with the BCG (see Figure~\ref{fig:Z7160_rgb}). To extract the clean components for these sources, we imaged again with \texttt{WSclean}, applying an inner \emph{uv}-cut of $5{\rm{k}}\uplambda$ (corresponding to a spatial scale of 41~arcsec or a linear scale of 158~kpc) to filter the diffuse emission of the mini-halo. This scale was chosen as it effectively suppressed the mini-halo without reducing sensitivity to compact sources within the boundary of the diffuse emission, allowing us to subtract the true contaminating compact source flux. The recovered clean components corresponding to both the BCG as well as sources in the vicinity of the mini-halo were then subtracted. Finally, we generated source-subtracted images using \texttt{WSclean} with the settings described in \S\ref{sec:postprocessing} to produce maps at two different resolutions: a `low-resolution' source-subtracted image at 15~arcsec resolution (using \texttt{robust}~$=-0.5$) and a `full-resolution' source-subtracted image at 8~arcsec resolution (using \texttt{robust}~$=-1.0$). In each case, we used a combination of appropriate \emph{uv}-tapering and subsequent image-plane smoothing to achieve the desired target resolution. We applied a Gaussian \emph{uv}-taper of 6 and 12~arcsec before convolving the final images to our target resolution of 8 and 15~arcsec, respectively.

\subsection{X-ray: \textit{Chandra}}
We reanalysed the archival \textit{Chandra} ObsID 4192 of MS~1455.0$+$2232 originally presented in \citet{Mazzotta2008}. The observation took place on 2003 September 5 (PI Mazzotta), is 95 ks long, and used \textit{Chandra} ACIS-I in VFAINT mode placing the cluster in the I3 chip. Data were processed from the \texttt{level=1} event file using CIAO v4.13 with CalDB v4.9.0. We searched for time periods affected by soft proton flares by inspecting the light curve extracted in the 0.5$-$7.0~keV band from the S2 chip that was kept on during the observation. The net time obtained after removing such periods is 88.6~ks. To maximise the signal-to-noise ratio from the ICM emission, exposure corrected images were created in the 0.5$-$2.0~keV energy band. Point sources were detected with the \texttt{wavdetect} task, visually inspected, and removed from the image by replacing them with random values extracted from neighbouring pixels. 

We produced projected temperature and entropy maps of the ICM performing spectral analysis in regions defined by \texttt{CONTBIN} v1.5 \citep{Sanders2006}. This algorithm allows us to bin regions of the image with similar surface brightness, and is particularly suited to studying the spatial variations of the thermodynamic quantities in galaxy clusters. During the binning process, we required that each spectral region had at least 1000 net counts in the 0.5$-$7.0 keV band. Thus, spectral extraction and fitting were performed with \texttt{XSPEC} v12.10.0c \citep{Arnaud1996}. Following \cite{Botteon2018b}, we accounted for the X-ray background by modelling the spectrum extracted from a region in the ACIS-I field-of-view free from the cluster emission, adopting an astrophysical and an instrumental component. The former is due to thermal emission from the galactic halo and local hot bubble plus the contribution of the unresolved emission from distant point sources, whereas the latter is due to the interaction of high-energy particles with the instrument, which was modelled using the analytical model of \cite{Bartalucci2014}. We adopted a thermal plasma to model the ICM emission, leaving the normalisation, metallicity, and temperature as free parameters. Projected values of entropy ($K$) were obtained combining the best fit temperature ($kT$) with the emission measure (EM) as $K = kT \times {\rm EM}^{-1/3}$. In the fitting, we adopted the abundance table from \cite{Asplund2009} and the column density $N_{\rm H} = 3.60 \times 10^{20}$ cm$^{-2}$ \citep{Willingale2013}.

\section{Results}\label{sec:results}

\begin{figure*}
\begin{center}
\includegraphics[width=0.95\linewidth]{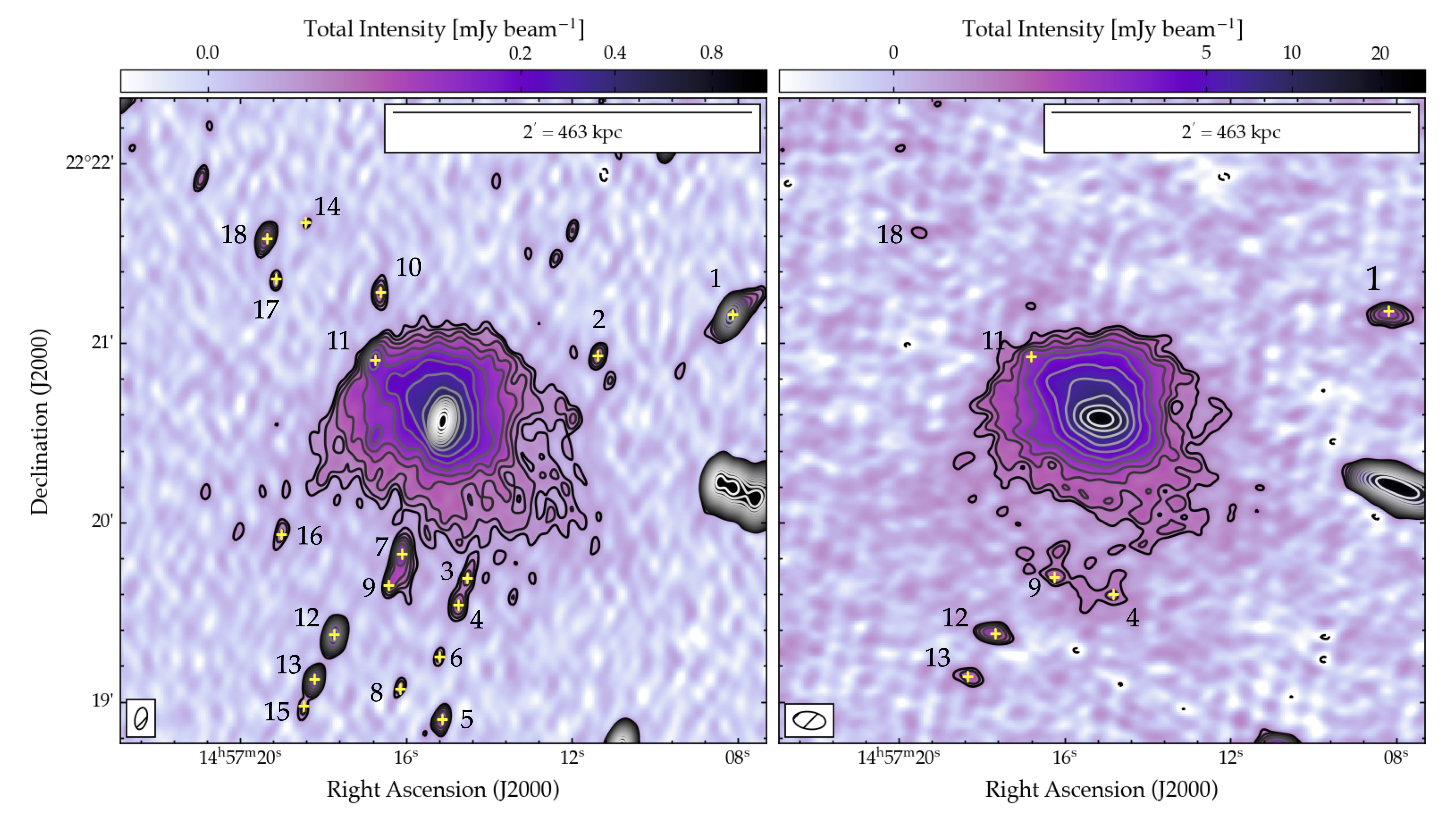}
\cprotect\caption{Radio continuum images of MS~1455.0$+$2232 with MeerKAT at 1283~MHz (\emph{left}) and LOFAR HBA at 145~MHz (\emph{right}) produced using \texttt{robust} $=-0.5$. Image properties (resolution and rms noise) are listed in Table~\ref{tab:img_summary}; the restoring beam is indicated by the hatched ellipse in the lower-left corner of each subplot. Contours start at $4\sigma$ and scale by a factor $\sqrt{2}$. The $-4\sigma$ contour is also shown by dashed lines. Yellow plus signs (`+') denote sources (other than the BCG) that were modelled and subtracted from each dataset; see Table~\ref{tab:measurement} for flux density measurements and cross-identifications.}
\label{fig:Z7160_continuum}
\end{center}
\end{figure*}

\subsection{Radio Continuum Properties}
Figure~\ref{fig:Z7160_continuum} presents our MeerKAT and LOFAR images of MS~1455.0$+$2232 prior to source subtraction, imaged using \texttt{robust}~$= -0.5$. The sensitivity and resolution of these images is presented in Table~\ref{tab:img_summary}. With the excellent sensitivity afforded by our observations, the mini-halo in MS~1455.0$+$2232 is recovered at high signal-to-noise at both 1283~MHz and 145~MHz. We also identify multiple compact and extended sources in the wider field, including a likely cluster-member head-tail radio galaxy to the West of the mini-halo (see also Figure~\ref{fig:Z7160_rgb}).

\begin{table}
\footnotesize
\centering
\caption{Summary of image properties. A `(sub)' indicates source-subtracted radio images. `LOFAR ILT' refers to our image made using the full International LOFAR Telescope. \label{tab:img_summary}}
\begin{tabular}{lcccr}
\hline
Telescope    &  \texttt{Robust} & RMS Noise                & Resolution  & PA       \\
 	   	     & &  [$\upmu$Jy beam$^{-1}$]  & [$\rm arcsec \times \rm arcsec$]    & [deg] \\
\hline\hline
\multirow{2}{*}{LOFAR}      & $-0.5$  & 140  & $10.8\times5.8$  & $+83$ \\
            & $-2.0$  & 829  & $3.9\times2.7$   & $+103$ \\
\hline
\multirow{2}{*}{LOFAR (sub)} & $-0.5$ & 159  & $15\times15$     & $+0$ \\
            & $-1.0$ & 179  & $8\times8$       & $+0$ \\
\hline
\multirow{2}{*}{MeerKAT}    & $-0.5$ & 5.9  & $7.4\times4.0$   & $+168$ \\
            & $-2.0$ & 19.8 & $5.5\times2.7$   & $+170$ \\
\hline
\multirow{2}{*}{MeerKAT (sub)} & $-0.5$ & 8.0  & $15\times15$     & $+0$ \\
            & $-1.0$ & 7.7  & $8\times8$      & $+0$ \\
                             
\hline
LOFAR ILT                   & $+0.5$ &  66.3 & $0.55\times0.30$ & $+6$  \\
\hline
\end{tabular}
\end{table}

As is visible in Figures~\ref{fig:Z7160_rgb} and \ref{fig:Z7160_continuum}, the mini-halo itself is asymmetric at both 1283~MHz and 145~MHz, appearing generally brighter to the north of the cluster centre (where we find the compact radio source associated with the BCG) with more extended fainter emission trailing toward the south and south-west. This fainter emission to the south and south-west has not been conclusively reported in previous radio studies at comparable resolution \citep[][]{Mazzotta2008,Venturi2008,Giacintucci2019} although we note a tentative detection at low resolution ($\sim15$ arcsec) by \cite{Venturi2008}.

However, as is visible in Figure~\ref{fig:Z7160_continuum}, there are multiple sources that lie in the vicinity of the mini-halo, or coincident with the diffuse emission. We aim to explore the diffuse radio emission from MS~1455.0$+$2232, and therefore excision of these contaminants is necessary. These were modelled and subtracted as discussed in Section~2.4.1; the sources that were subtracted (aside from the BCG) are identified by yellow plus-signs (`+') in Figure~\ref{fig:Z7160_continuum}. Flux density measurements for these sources, along with SDSS cross-identifications and redshifts, are presented in Table~\ref{tab:measurement}.

\subsubsection{Source-Subtracted Images}
\begin{figure*}
\begin{center}
\includegraphics[width=0.82\linewidth]{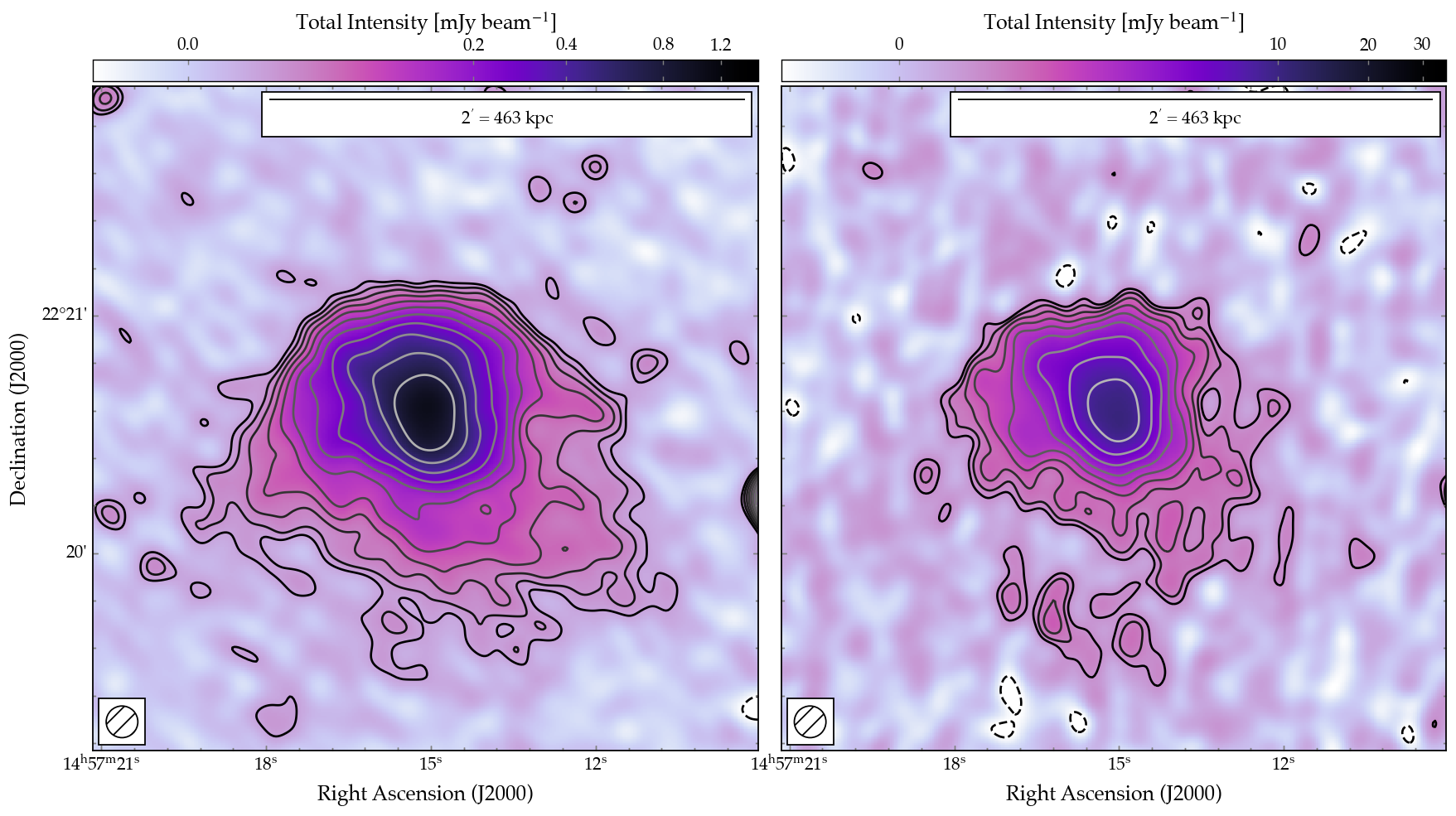} \\
\includegraphics[width=0.82\linewidth]{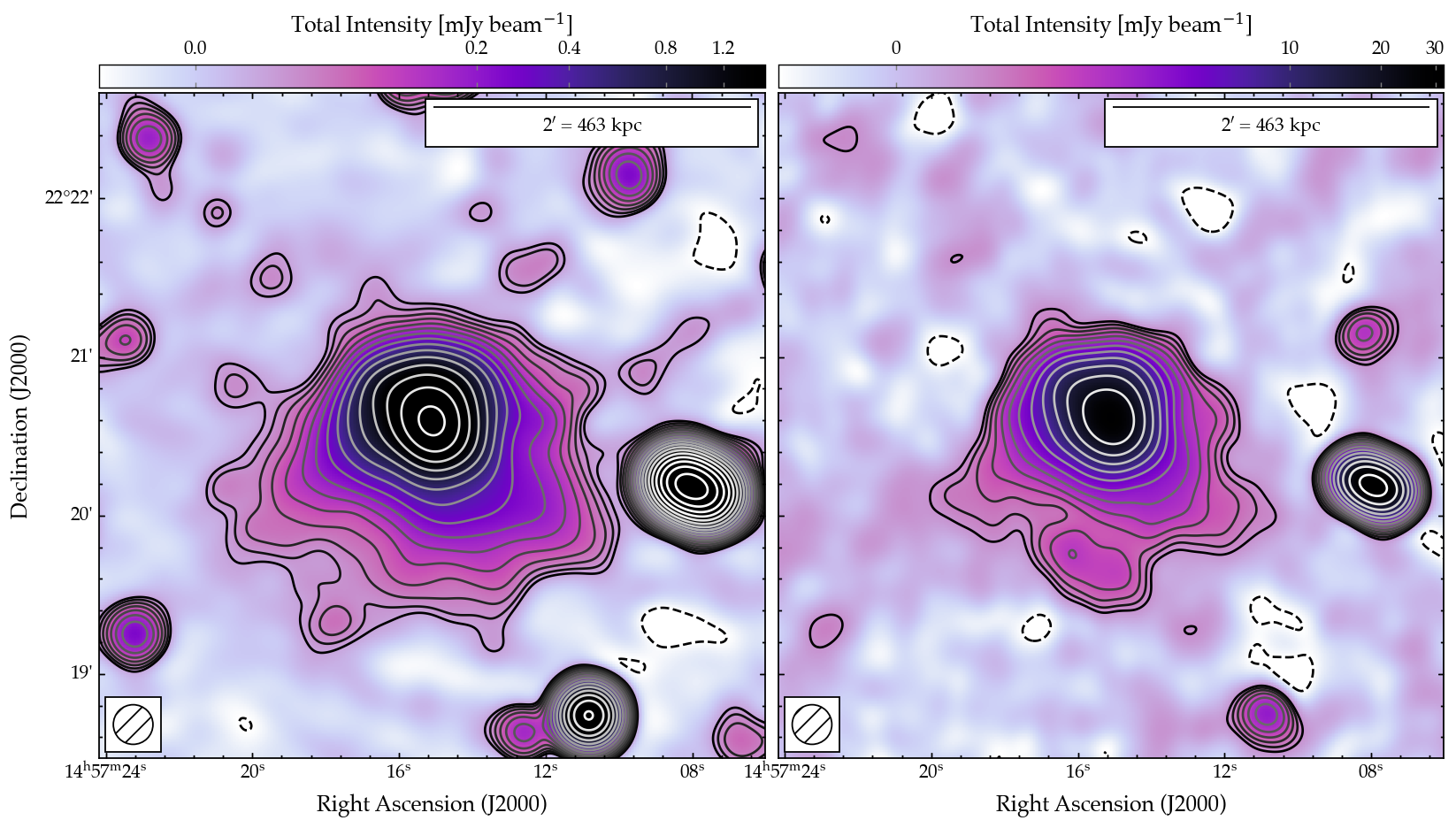}
\cprotect\caption{Source-subtracted images of MS~1455.0$+$2232 at 1283~MHz (MeerKAT; \emph{left}) and 145~MHz (LOFAR; \emph{right}), shown at 8~arcsec resolution (\emph{top}) and 15~arcsec resolution (\emph{bottom}). Contours are shown at intervals of $\sqrt{2}$ starting from $3\sigma$, where $\sigma = 7.3~\upmu$Jy~beam$^{-1}$ at 1283~MHz and $159~\upmu$Jy~beam$^{-1}$ at 145~MHz; the $-3\sigma$ contour is shown as a dashed line. The colour scale ranges from $-3\sigma$ to $200\sigma$ and is shown on an arcsinh stretch to emphasise diffuse emission. The scale bar denotes the linear size at $z=0.258$ corresponding to an angular scale of 2~arcmin.}
\label{fig:Z7160_subtracted}
\end{center}
\end{figure*}

Figure~\ref{fig:Z7160_subtracted} presents our source-subtracted, \emph{uv}-tapered images of MS~1455.0$+$2232 at 8~arcsec resolution (\emph{top row}) and 15~arcsec resolution (\emph{bottom row}). With the enhanced sensitivity to very low surface brightness radio emission, the diffuse radio emission of the mini-halo is detected to a much greater extent. From our 15~arcsec resolution MeerKAT image at 1283~MHz, we measure a largest angular scale of 152~arcsec, corresponding to a largest linear size (LLS) of 586~kpc with our cosmology. This is 69 per cent larger than previously reported by \cite{Venturi2008} at 610~MHz, who measured a LAS (LLS) of 90~arcsec (347~kpc with our cosmology). At 145~MHz, the LAS is 122~arcsec (471~kpc).

We note that it is unusual to find non-thermal cluster phenomena that are more extended at higher frequencies than lower frequencies. The faint, steep-spectrum radio sources often associated with clusters typically trace weak and/or inefficient (re-)acceleration processes in the ICM, and are more commonly found to decrease in size with increasing frequency \citep[reflecting the nature of the acceleration mechanism, e.g.][]{Rajpurohit2021b,Rajpurohit2021c}. However, our MeerKAT images achieve a very low rms noise of $7.3~\upmu$Jy~beam$^{-1}$ at 1283~MHz; as such, a source that has a $1\sigma$ flux density at each frequency would have a spectral index of $\alpha = -1.37$. This means that the larger extent at higher frequencies does not necessarily imply a spectral flattening.

\subsection{X-ray Properties}
\begin{figure*}
\begin{center}
\includegraphics[width=0.87\linewidth]{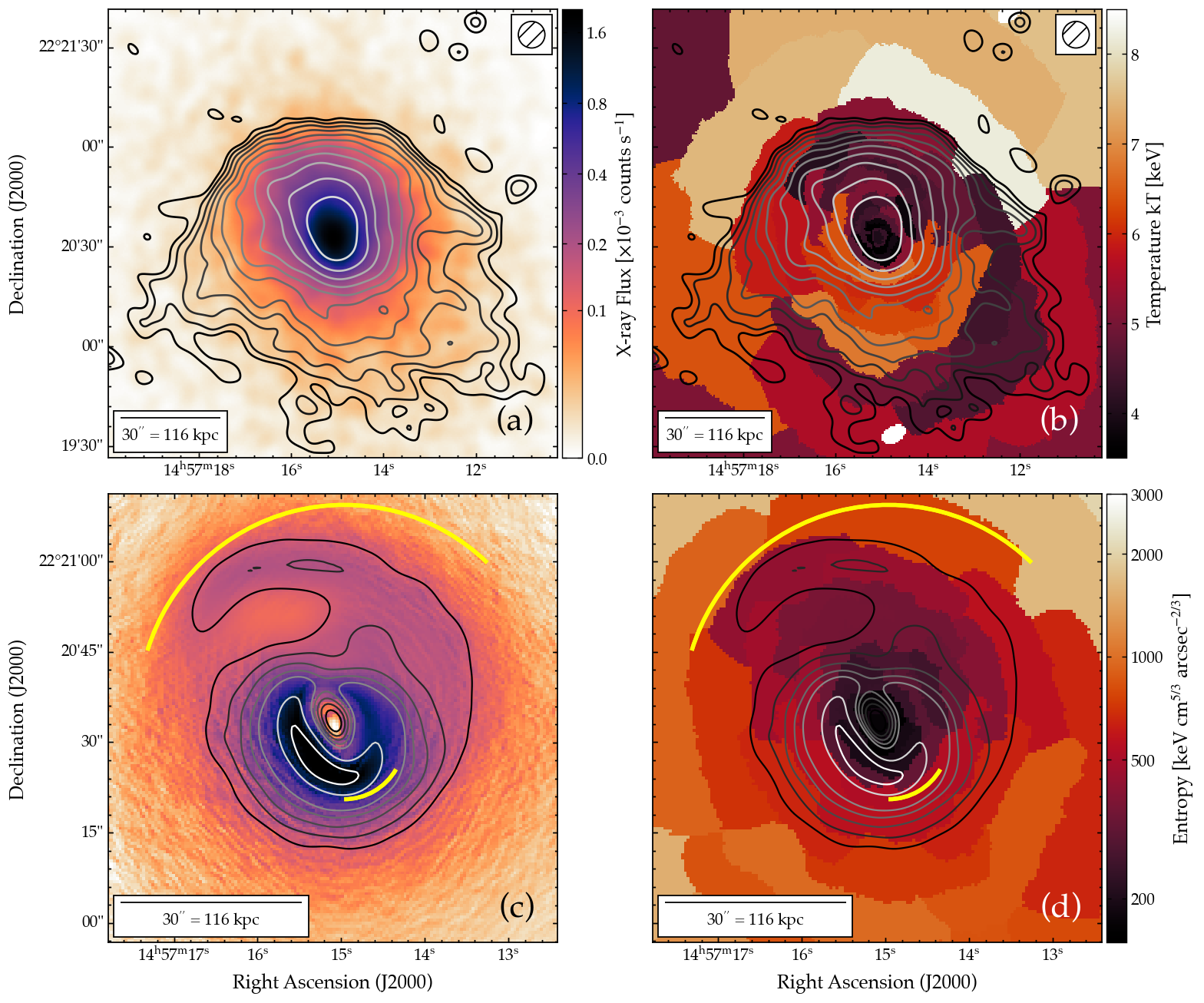}
\cprotect\caption{Multi-wavelength images of MS~1455.0$+$2232. Background colormaps show different quantities derived from our \textit{Chandra} data: (a) 0.5$-$2.0~keV X-ray flux, smoothed with a Gaussian of 8~arcsec FWHM, (b) ICM temperature, (c) X-ray surface brightness gradient, (d) ICM entropy. In panels (a) and (b), contours show 1283~MHz surface brightness measured by MeerKAT at 8~arcsec resolution as per Figure~\ref{fig:Z7160_subtracted} (the resolution is shown by the hatched circle in the upper-right corner). Contours in panels (c) and (d) denote the X-ray gradient, smoothed with a Gaussian of FWHM 3~pixels, with levels chosen to highlight the structure of the sloshing spiral. Yellow arcs trace the cold fronts identified by \protect\cite{Mazzotta2001}; our new analysis reveals that these trace features visible in the larger-scale sloshing spiral. The bright `hotspot' at the centre of the sloshing spiral is clearly visible at the centre of panel (c).}
\label{fig:xray_components}
\end{center}
\end{figure*}

Figure~\ref{fig:xray_components} shows the X-ray surface brightness measured by \textit{Chandra} in the 0.5$-$2.0~keV band, as well as several quantities derived from our X-ray data. These are the ICM temperature (panel b), X-ray surface brightness gradient (panel c), and ICM entropy (panel d).

The X-ray gradient map is particularly useful to highlight the surface brightness discontinuities in the ICM, such as shocks and cold fronts, and study their connection with the radio emission. The map shown Figure~\ref{fig:xray_components} panel (c) was obtained using the adaptively smoothed Gaussian Gradient Magnitude (GGM) filter recently introduced by \citet{Sanders2021}. This method adds in quadrature the gradients computed along the \textit{x}- and \textit{y}-axes from the $\log_{10}$ value of an adaptively-smoothed X-ray image of the cluster (for MS~1455.0$+$2232, we adopted a smoothing signal-to-noise ratio of 25). Our image clearly shows a spiral-like structure that is often associated with core-sloshing in relaxed clusters: a sloshing spiral \citep[see for example][]{Sanders2016a,Sanders2016b,Werner2016,Douglass2018}. 

The sloshing spiral is asymmetric around its peak (which is traced by the `hotspot' in Figure~\ref{fig:xray_components}, panel c): it occupies an elliptical region 66~arcsec~$\times$~60~arcsec (at PA $\sim29\degree$) centred on (J2000 RA,~Dec) (14:57:15.14,~$+$22:20:37.6). This corresponds to a physical scale of 254~kpc~$\times$~232~kpc. Our new analysis reveals that the cold fronts previously identified by \cite{Mazzotta2008} form part of the larger spiral structure in MS~1455.0$+$2232; this kind of structure almost perfectly matches prototypical core-sloshing spirals seen in numerical simulations \citep[e.g.][]{Ascasibar2006,ZuHone2013}.

Panel (d) of Figure~\ref{fig:xray_components} shows the (projected) entropy derived from our \textit{Chandra} data. We highlight that the low-entropy regions seen in panel (d) closely follow the boundaries of the sloshing spiral, suggesting that low-entropy gas is being disturbed by the core sloshing. Indeed, these kind of structures have been observed in sloshing clusters \citep[][]{Fabian2006,Ghizzardi2014} and are also predicted by numerical simulations \citep[][]{Ascasibar2006,ZuHone2013}. The simulations by \citeauthor{Ascasibar2006} in particular suggest that this kind of sloshing structure can be generated by minor or off-axis merger events. While there is no evidence of substructure in the galaxy distribution of MS~1455.0$+$2232 that might indicate a minor merger, to the best of our knowledge such an investigation has not been performed to-date, and is beyond the scope of this paper.

\begin{figure*}
\begin{center}
\includegraphics[width=0.9\linewidth]{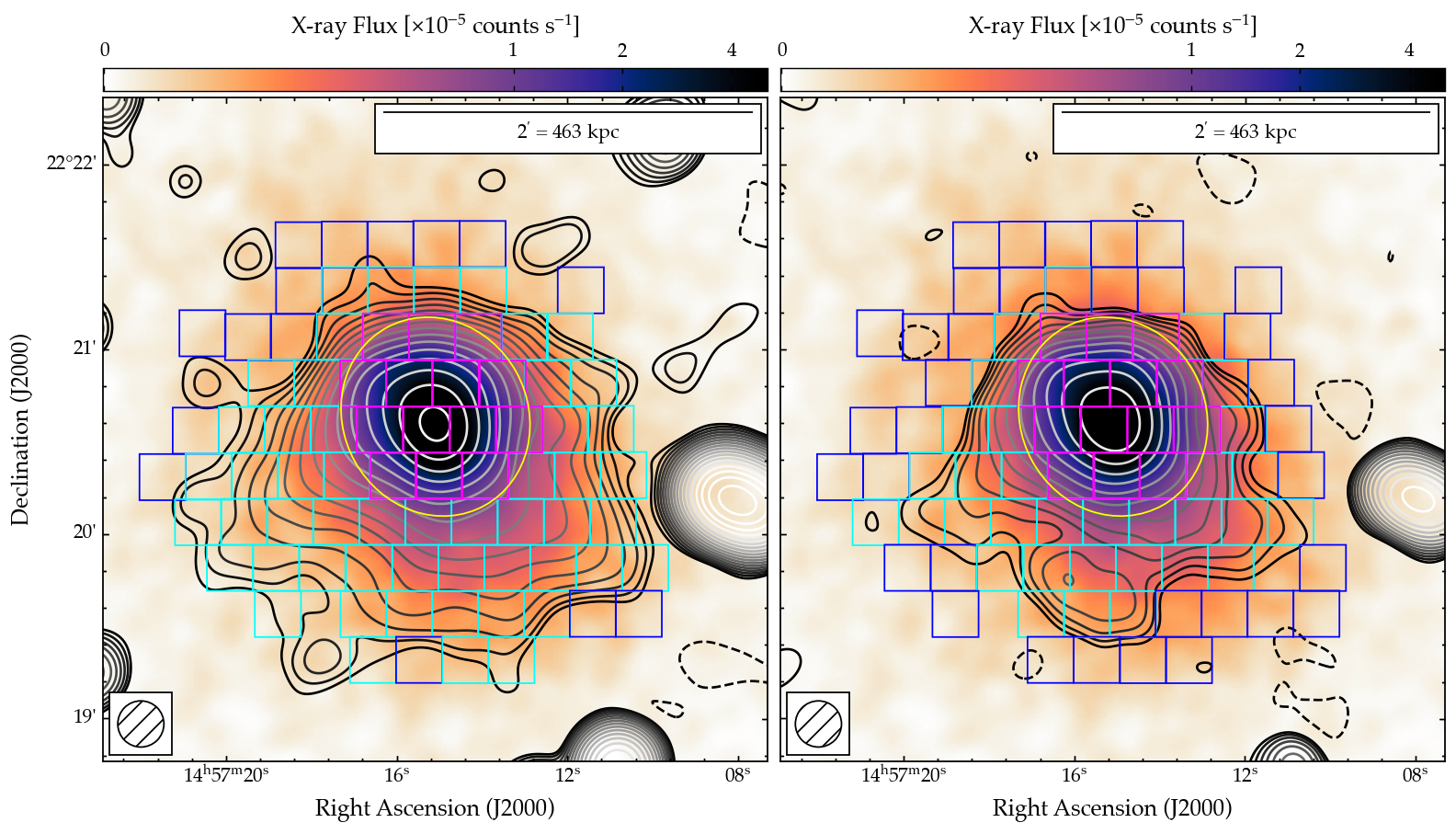}
\caption{Multi-wavelength images of the mini-halo in MS~1455.0$+$2232. Colour map shows the X-ray surface brightness measured by \textit{Chandra} in the 0.5$-$2~keV band, smoothed with a 15~arcsec FWHM Gaussian to match the resolution of our radio data. Contours denote the source-subtracted radio data (\emph{left}: MeerKAT; \emph{right}: LOFAR) at 15~arcsec resolution, starting at $3\sigma$ and scale by $\sqrt{2}$, as per Figure~\ref{fig:Z7160_subtracted}. Boxes show the regions used to profile the radio/X-ray correlations, with magenta boxes indicating regions within the boundary of the sloshing spiral (identified by the yellow ellipse), cyan boxes indicating regions outside the sloshing spiral, and blue boxes indicating regions where the average radio surface brightness is below the corresponding $3\sigma$ level.}
\label{fig:radio_xray_overlay}
\end{center}
\end{figure*}

Figure~\ref{fig:radio_xray_overlay} presents multi-wavelength images of MS~1455.0$+$2232, showing the X-ray surface brightness measured by \textit{Chandra} (smoothed with a Gaussian of 15~arcsec FWHM) and radio contours from our source-subtracted MeerKAT and LOFAR data at 15~arcsec resolution overlaid to guide the reader. We also trace the outer edge of the sloshing spiral identified in Figure~\ref{fig:xray_components}. Finally, Figure~\ref{fig:radio_xray_overlay} also shows boxes which will be later used to study point-to-point correlations between thermal and non-thermal properties, as well as profiling the spectral index.

From Figure~\ref{fig:radio_xray_overlay}, the mini-halo in MS~1455.0$+$2232 extends far beyond the boundaries of the sloshing spiral: the diffuse synchrotron emission fills almost the entire X-ray-emitting region (above a count rate of $5\times10^{-7}$ counts s$^{-1}$). To the north, the mini-halo extends for 20~arcsec (77~kpc) beyond the boundary of the sloshing spiral. However, to the south-east, south, and south-west, the mini-halo extends 56~arcsec (or 216~kpc) beyond the outer edge of the sloshing spiral, as traced by our MeerKAT data at 1283~MHz.

Many historic mini-halo studies have not detected significant radio emission outside of the sloshing regions in relaxed cool-core clusters. As such, most theoretical models that associate mini-haloes with turbulence predict that particle acceleration (and thus diffuse synchrotron emission) will be limited to the relatively small scales of the sloshing region \citep[][]{ZuHone2013}. MS~1455.0$+$2232 has historically been one of the prototypical sloshing clusters where the mini-halo is believed to be generated by turbulence induced by core sloshing \citep{Mazzotta2008}, and thus the extent of the sloshing spiral and the diffuse mini-halo would appear to be in tension.


\subsection{Radio Emission from the Brightest Cluster Galaxy}

\subsubsection{Integrated Spectral Energy Distribution}
To measure the flux density of the radio counterpart to the BCG, we used our uniform-weighted MeerKAT and LOFAR images with an inner \emph{uv}-cut of $5~\rm k \lambda$. The BCG radio core remains unresolved at the resolution limit of these data (i.e. the core size is less than $2.6$~arcsec or a physical scale of 10~kpc). We measure an integrated flux density of $S_{\rm 1283~MHz} = 4.74 \pm 0.24$~mJy from our MeerKAT image, and $S_{\rm 145~MHz} = 11.93 \pm 1.24$~mJy for the compact radio component detected in our uniform-weighted (\texttt{robust}~$=-2$) LOFAR Dutch array map.

Thus, we find a two-point spectral index $\alpha_{\rm 1283~MHz}^{\rm 145~MHz} = -0.42 \pm 0.11$. This is significantly flatter than the spectral index previously reported by \cite{Giacintucci2019}, who found $\alpha_{\rm 1.4~GHz}^{\rm 610~MHz} = -0.85 \pm 0.11$ (accounting for differences in spectral index convention). However, this previously-reported spectral index is wholly tied to the calibration and deconvolution precision of both datasets.

\begin{table}
\centering
\caption{Flux density measurements for the unresolved radio core of the BCG in MS~1455.0$+$2232. Note that all measurements are converted to the \protect\cite{ScaifeHeald2012} flux density scale, aside from the 28.5~GHz measurement which is quoted directly from \protect\cite{Coble2007}. \label{tab:core_fluxes}}
\begin{tabular}{lrr}
\hline
Frequency    & Flux density    \\
{} [GHz]       & [mJy]             \\
\hline\hline
0.145        & $11.93 \pm 1.24$ \\
0.610        & $6.26 \pm 0.36$  \\
1.283        & $4.74 \pm 0.24$  \\
1.4          & $4.56 \pm 0.30$  \\
3.0          & $2.95 \pm 0.32$  \\
4.85         & $2.27 \pm 0.12$  \\
8.46         & $1.41 \pm 0.08$  \\
28.5         & $0.96 \pm 0.10$  \\
\hline
\end{tabular}
\end{table}

\begin{figure}
\begin{center}
\includegraphics[width=0.95\linewidth]{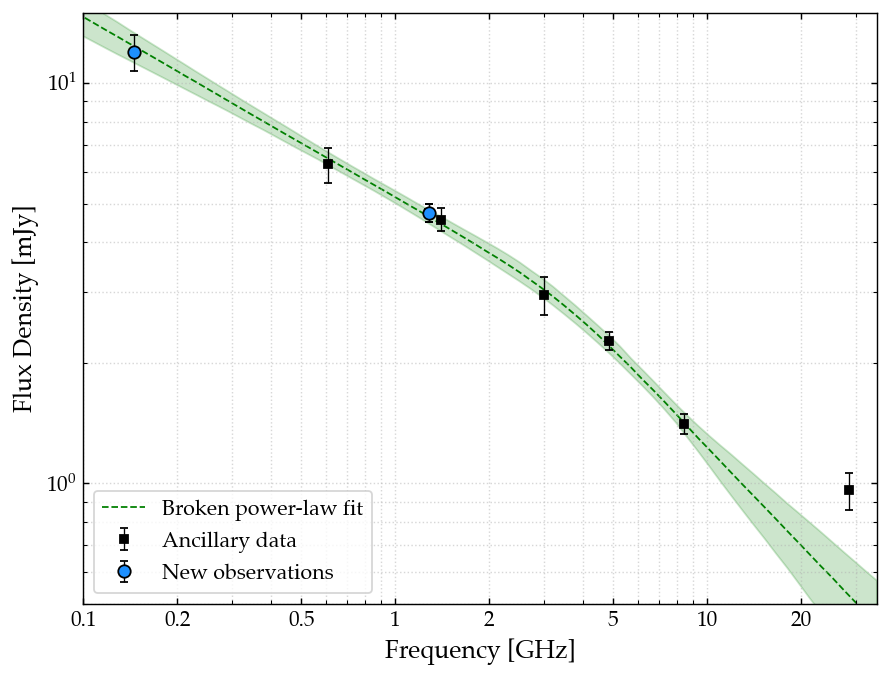}
\cprotect\caption{Radio spectral energy distribution (SED) for the compact radio source associated with the brightest cluster galaxy in MS~1455.0$+$2232. Our new MeerKAT and LOFAR measurements at 1283~MHz and 145~MHz are shown as blue circles; measurements from reprocessed archival data are shown as black squares. Dashed green line denotes our best-fit broken power-law spectrum, with $\alpha_{\rm low} = -0.45 \pm 0.05$ below the break frequency ($\nu_{\rm b} ~ = ~ 3.12 ~ \pm ~ 1.35$~GHz) and $\alpha_{\rm high} = -0.81 \pm 0.18$ above the break; the green shaded region indicates the $1\sigma$ uncertainty explored by \textsc{emcee}. Note that the 28.5~GHz measurement was excluded from our fitting due to potential source variability at this frequency.}
\label{fig:BCG_SED}
\end{center}
\end{figure}

Alongside our new MeerKAT and LOFAR observations, our reprocessing of archival narrow-band datasets provides the broad-band coverage necessary to explore the radio spectral energy distribution (SED) of the BCG. Table~\ref{tab:core_fluxes} lists the measurements from our new MeerKAT and LOFAR observations as well as from our reprocessed ancillary data, plus a 28.5~GHz measurement with the Berkeley-Illinois-Maryland Association (BIMA) interferometer which is quoted directly from \cite{Coble2007}. Figure~\ref{fig:BCG_SED} presents the result of our analysis. The SED is well-characterised by a relatively flat-spectrum power-law between 145~MHz and 1.4~GHz, and exhibits departure from this power-law behaviour somewhere in the range 1.4~GHz to 3~GHz. We find a 610~MHz flux density of $S_{\rm 610~MHz} = 6.26 \pm 0.63$~mJy, measured from our final image produced by the \texttt{SPAM} pipeline. This is around 34 per cent lower than the value previously reported by \cite{Mazzotta2008}; however, as is visible from Figure~\ref{fig:BCG_SED} our newly-measured 610~MHz flux density is entirely consistent with the overall SED profile.

We modelled the behaviour of the SED using a broken power-law fit, and explored the uncertainty region using an `affine invariant' Markov-chain Monte Carlo (MCMC) ensemble sampler \citep{Goodman2010} as implemented by the \textsc{emcee} package \citep{emcee} to constrain the model. We note that the 28.5~GHz measurement was omitted from our fitting routine; the BIMA observations reported by \cite{Coble2007} were performed in the 1996--2002 period, and at these frequencies, variability on year-decade timescales is likely. More recent observations at comparable frequency and resolution would be required to quantify this, however.

Our fitting routine yielded a best-fit broad-band spectral index $\alpha_{\rm{low}} = -0.45 \pm 0.05$ below the break frequency, which is consistent with the two-point spectral index derived earlier. Above the break frequency, we find a best-fit spectral index $\alpha_{\rm{high}} = -0.81 \pm 0.18$. The break frequency itself is poorly-constrained by the available data: we find $\nu_{\rm b} = 3.1 \pm 1.4$~GHz. Further broad-band data from (for example) the VLA would be required to more accurately pinpoint the break frequency for this radio source. 

Given our spectral index convention, the $k$-corrected radio power is given by the following equation:
\begin{equation}\label{eq:radio_lum}
    P_{\nu} = 4 \pi \,  D_{\rm L}^2 \, S_{\nu} \, (1 + z)^{-(1 + \alpha)} \quad [\rm{W ~ Hz}^{-1}] 
\end{equation}
where $D_{\rm L} = 1259.7$~Mpc is the luminosity distance at the cluster redshift of $z = 0.258$, and $S_{\nu}$ is the flux density at frequency $\nu$. From our broad-band SED fit, we find $S_{\rm 1.4~GHz} = 4.45 \pm 0.18$~mJy. Thus, Equation~\ref{eq:radio_lum} yields a radio luminosity of ${\rm P}_{\rm 1.4~GHz} ~ (\rm BCG) = (7.44 \pm 0.30) \times 10^{23}$~W~Hz$^{-1}$. 


This radio power is a factor 3.5 lower than the `BCG steep' radio power at 1~GHz derived by \cite{RichardLaferriere2020} from flux density measurements presented by \cite{Hogan2015}. However, as noted in the supplementary material of \cite{Hogan2015}, their 1.4~GHz measurement likely recovers `more diffuse emission from the non-core component', which may reflect a partial detection of the mini-halo. As such, the radio power derived from our measurements is not directly comparable with previous values reported by \cite{Hogan2015} and \cite{RichardLaferriere2020}.

\subsubsection{Sub-arcsecond LOFAR Images}
\begin{figure}
\begin{center}
\includegraphics[width=0.95\linewidth]{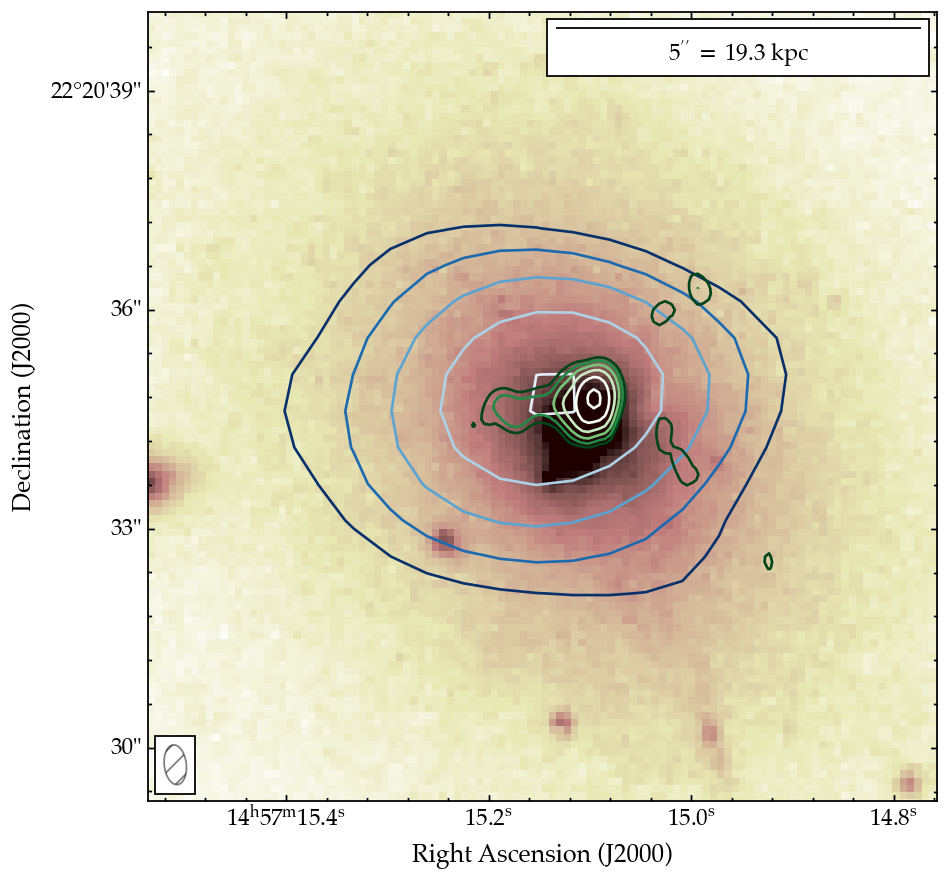}
\cprotect\caption{Multi-wavelength zoom on the BCG of MS~1455.0$+$2232. The background colormap shows a \emph{Hubble Space Telescope} image taken with the Wide-Field and Planetary Camera 2 (WFPC2) in the $F606W$ filter. Blue contours show the \texttt{uniform}-weighted (\texttt{robust}~$-2$) LOFAR surface brightness at a resolution of $3.9\times2.7$~arcsec$~\times$~arcsec; the first contour is at $4\sigma$. Green contours show the radio surface brightness from our full ILT image at a resolution of $0.55\times0.30$~arcsec~$\times$~arcsec (indicated by the hatched ellipse); the first contour is at $5\sigma$. All contours increment by a factor of $\sqrt{2}$.}
\label{fig:LOFAR_ILT}
\end{center}
\end{figure}

When measuring the integrated flux of the mini-halo, the dominant source of contamination is the compact radio source associated with the BCG. Figure~\ref{fig:LOFAR_ILT} presents a zoom on the BCG of MS~1455.0$+$2232, overlaid with contours from our high-resolution LOFAR image (produced using the Dutch array) as well as our sub-arcsecond resolution image produced using the full ILT.

From Figure~\ref{fig:LOFAR_ILT}, we see that the radio source associated with the BCG of MS~1455.0$+$2232 remains compact at a resolution of the Dutch array (around $2.5$~arcsec). It is only with the sub-arcsecond imaging capability of the ILT that we are able to decompose this compact radio source into multiple components: a dominant western component that appears to be slightly misaligned with the BCG, and a much weaker component that extends to the east.

We measure a total flux density of $S_{\rm 145~MHz} = 11.93 \pm 1.24$~mJy for the compact radio component detected in our high-resolution LOFAR Dutch array map. From the ILT map, the primary (western) component presents as a marginally-resolved radio source with a peak flux density $S_{\rm 145~MHz} = 1.72$~mJy beam$^{-1}$. Integrating above the $4\sigma$ contour of our ILT map, we find a total integrated flux density of $S_{\rm 145~MHz} = 6.77 \pm 0.70$~mJy for both components. As such, our ILT image is around $57\%$ flux-complete; a significant fraction of the (unresolved) flux of the radio core lies below the surface brightness sensitivity level of our present ILT image. 

This relatively poor flux completeness of our ILT image, combined with the lack of higher-frequency maps capable of achieving the same resolution, means that we cannot yet determine the nature of these multiple components. Deeper observations with the ILT would increase our flux completeness and allow us to better understand the morphology of this source. Spectral information would be required to better understand the nature of these components, with further observations at complementary frequencies using instruments such as eMERLIN and/or lower-frequency LOFAR Low-Band Antenna (LBA) VLBI involving the NenuFAR station in superstation mode \citep[e.g.][]{nenufar,2020A&A...637A..51B}.

\subsubsection{Radiative Age}
The spectral behaviour of resolved radio galaxies can be used to determine their magnetic field and radiative age assuming that the energy budget is equally distributed between magnetic fields and relativistic particles.

Following Equations~24 and 26 of \cite{Govoni2004}, we derive the equipartition magnetic field by modelling the two components of the BCG as two spheres of radius 0.6~arcsec and 0.35~arcsec (as seen in our ILT map in Figure~\ref{fig:LOFAR_ILT}), obtaining a volume of $\sim62$~kpc$^3$. We assume a filling factor $\Phi = 1$ and a proton-to-electron energy ratio $k=1$. The synchrotron luminosity was derived from the flux density measured at 145~MHz in our ILT map. The BCG equipartition magnetic field is $B_{\rm eq} = 1~\upmu$G.

Assuming the continuous injection model \citep[CI;][]{pacholczyk}, where freshly injected particles compensate the radiative losses, it is possible to derive the radiative age of the source ($\tau_{\rm s}$) as:
\begin{equation}
\tau_{\rm s} = 1590 \frac{B^{0.5}}{(B^2 + B_{\rm IC}^2) ~ [(1+z)\nu_{\rm b}]^{0.5}} \quad [\rm{Myr}] \, , 
 \label{eq:ts}
\end{equation}
where $B_{\rm IC}$ is the equivalent magnetic field due to inverse Compton losses that depends on redshift as $B_{\rm IC} = 3.25 (1+z)^2~\upmu$G, $B$ is the BCG magnetic field, while $\nu_{\rm b} = 3.1 \pm 1.4$~GHz is the break frequency of the SED. Taking the BCG magnetic field to be the equipartition magnetic field derived earlier (i.e. $B = B_{\rm eq} = 1~\upmu$G), we obtain a radiative age of $\tau_{\rm s} \simeq 46 \pm 20$~Myr. We note that the dominant contribution to this uncertainty budget is the significant uncertainty in the break frequency.

\section{Analysis: The Mini-Halo in MS~1455.0+2232}\label{sec:analysis}

\subsection{Spectral Properties}

\subsubsection{Integrated Spectral Properties}
Integrating above the $3\sigma$ contour in the corresponding MeerKAT and LOFAR images of the mini-halo in Figure~\ref{fig:Z7160_subtracted}, we find a total flux density recovered by MeerKAT of $S_{\rm 1283~MHz} = 18.7 \pm 0.9$~mJy. For LOFAR, we measure a total integrated flux density of $S_{\rm 145~MHz} = 154.5 \pm 15.6$~mJy. Thus, our integrated spectral index for the mini-halo in MS~1455.0$+$2232 is $\alpha^{\rm 145~MHz}_{\rm 1283~MHz} = -0.97 \pm 0.05$.

This is significantly flatter than was previously reported by \cite{Giacintucci2019} for this cluster: $\alpha^{\rm 610~MHz}_{\rm 1.4~GHz} = -1.46 \pm 0.22$. The cause of this difference is readily apparent when comparing our Figure~\ref{fig:Z7160_subtracted} with their Figure~15. Only the brightest central region of the mini-halo is recovered by the VLA C-configuration observations presented by \citeauthor{Giacintucci2019}; as such they recover an integrated 1.4~GHz flux density $S_{\rm 1.4~GHz} = 8.5 \pm 1.1$~mJy. Our deeper MeerKAT observations are far more sensitive to the low-surface-brightness regions of the mini-halo, and as such our integrated flux density is more than double the value recovered by \citeauthor{Giacintucci2019}.

Few well-studied examples of mini-halo spectra exist in the literature; however, the integrated spectral index we find for MS~1455.0$+$2232 is significantly flatter than the mini-haloes hosted by RX~J1532.9$+$3021 ($\alpha = -1.20 \pm 0.03$), the Perseus cluster ($\alpha = -1.21 \pm 0.05$) and the Ophiuchus cluster ($\alpha = -1.56 \pm 0.04$) for example \citep{Giacintucci2014b}. 

On the other hand, our integrated spectral index is consistent with some known mini-haloes. The mini-haloes in Abell~2667 \citep[$\alpha = -1.0\pm0.2;$][]{Giacintucci2019}, the Phoenix Cluster \citep[$\alpha = -0.95\pm0.10;$][]{Raja2020,Timmerman2021} and RX~J1720.1$+$2638 \citep[$\alpha = -0.99 \pm 0.03$;][]{Biava2021} all show a similar integrated spectral index to MS~1455.0$+$2232. Deep observations at complementary frequencies would be required to explore the spectral properties of this mini-halo further. However, to search for the presence of a spectral break, higher-frequency observations would be required with instruments such as the VLA (S- and/or C-bands) or the MeerKAT S-band receiver system. Such observations could provide important evidence when attempting to differentiate between the particle acceleration mechanism at work in MS~1455.0$+$2232.

\subsubsection{Mini-Halo Luminosity}
Using our 1283~MHz integrated flux density measurement and our integrated spectral index, we can use Equation~\ref{eq:radio_lum} to estimate the mini-halo luminosity. Using the canonical cluster redshift of $z = 0.258$, we find ${\rm P}_{\rm 1.4~GHz} \, ({\rm MH}) = (3.24 \pm 0.16) \times 10^{24} ~ {\rm W} ~ {\rm Hz}^{-1}$. Due to differences in the methodology and the measured spectral index, this figure is not directly comparable to the radio power measurements previously reported by \cite{Venturi2008} and \cite{Giacintucci2019}. In any case, despite deriving a larger value, our radio power measurement is consistent with the broader mini-halo population, which typically exhibit ${\rm P}_{\rm 1.4~GHz} \, (\rm MH) \sim 10^{23} - 10^{25}~ {\rm W} ~ {\rm Hz}^{-1}$ \citep[][and references therein]{vanWeeren2019}. 

\begin{figure}
\begin{center}
\includegraphics[width=0.95\linewidth]{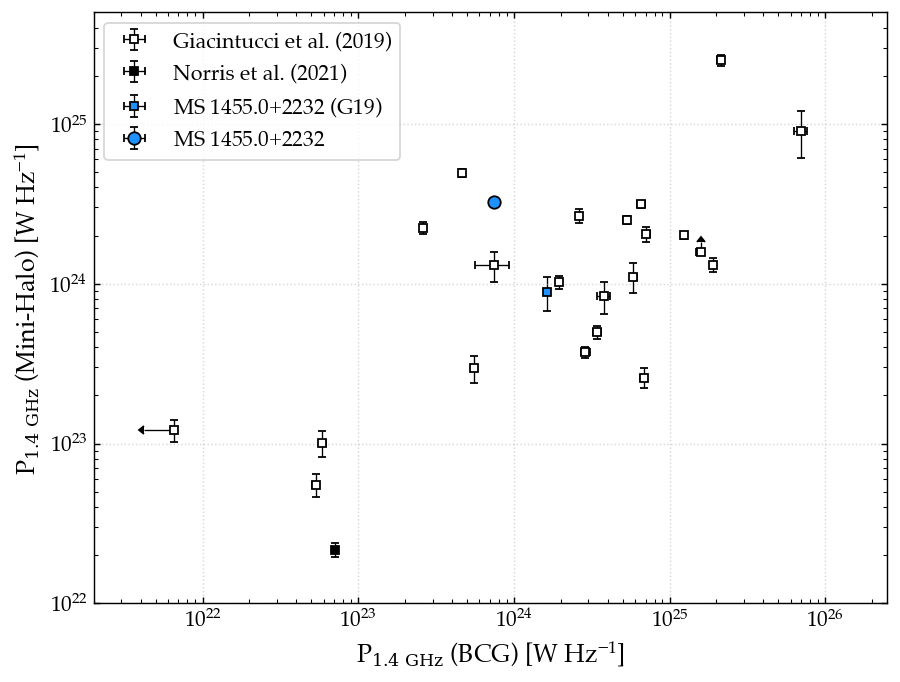}
\cprotect\caption{Power scaling plane for mini-haloes, showing how the mini-halo radio power, ${\rm P}_{\rm 1.4~GHz} \, (\rm MH)$, scales with BCG radio power, ${\rm P}_{\rm 1.4~GHz} \, (\rm BCG)$. Open squares denote known mini-haloes from the sample of \cite{Giacintucci2019}, adjusted for differences in assumed cosmology; the black square denotes the mini-halo candidate recently reported by \cite{Norris2021}. Blue points indicate the position of MS~1455.0$+$2232, with the square denoting the measurements from \cite{Giacintucci2019} and the circle indicating our new measurements (note that the uncertainties on our new measurements are not visible at this scale).}
\label{fig:mh_scaling}
\end{center}
\end{figure}

When combined with the BCG radio power derived earlier --- ${\rm P}_{\rm 1.4~GHz} \, (\rm BCG) = (7.44 \pm 0.30) \times 10^{23}$~W~Hz$^{-1}$ --- we can examine where MS~1455.0$+$2232 now lies in known power scaling planes. Figure~\ref{fig:mh_scaling} presents the ${\rm P}_{\rm 1.4~GHz} \, (\rm BCG)$/${\rm P}_{\rm 1.4~GHz} \, (\rm MH)$ plane for known radio mini-haloes, using the sample presented by \cite{Giacintucci2019}. We also add the mini-halo candidate recently reported by \cite{Norris2021} at 943~MHz in the Evolutionary Map of the Universe \citep[EMU;][]{Norris2011} Pilot Survey region, with measurements scaled to 1.4~GHz using canonical spectral index values $\alpha = -0.7$ for the BCG and $\alpha = -1$ for the mini-halo.

From Figure~\ref{fig:mh_scaling}, we see that our new results place the mini-halo in MS~1455.0$+$2232 toward the middle range of the scaling plane between mini-halo radio power and BCG radio power for the known population. However, we also see that our new measurements of the mini-halo and BCG radio power for this cluster have significantly shifted the position of MS~1455.0$+$2232 in this scaling plane, compared to the values reported by \cite{Giacintucci2019}. We will revisit the scaling relation between ${\rm P}_{\rm 1.4~GHz} \, (\rm MH)$ and ${\rm P}_{\rm 1.4~GHz}\,(\rm BCG)$ in future for the full sample from our homogeneously-selected mini-halo census.

\begin{figure*}
\begin{center}
\includegraphics[width=0.85\linewidth]{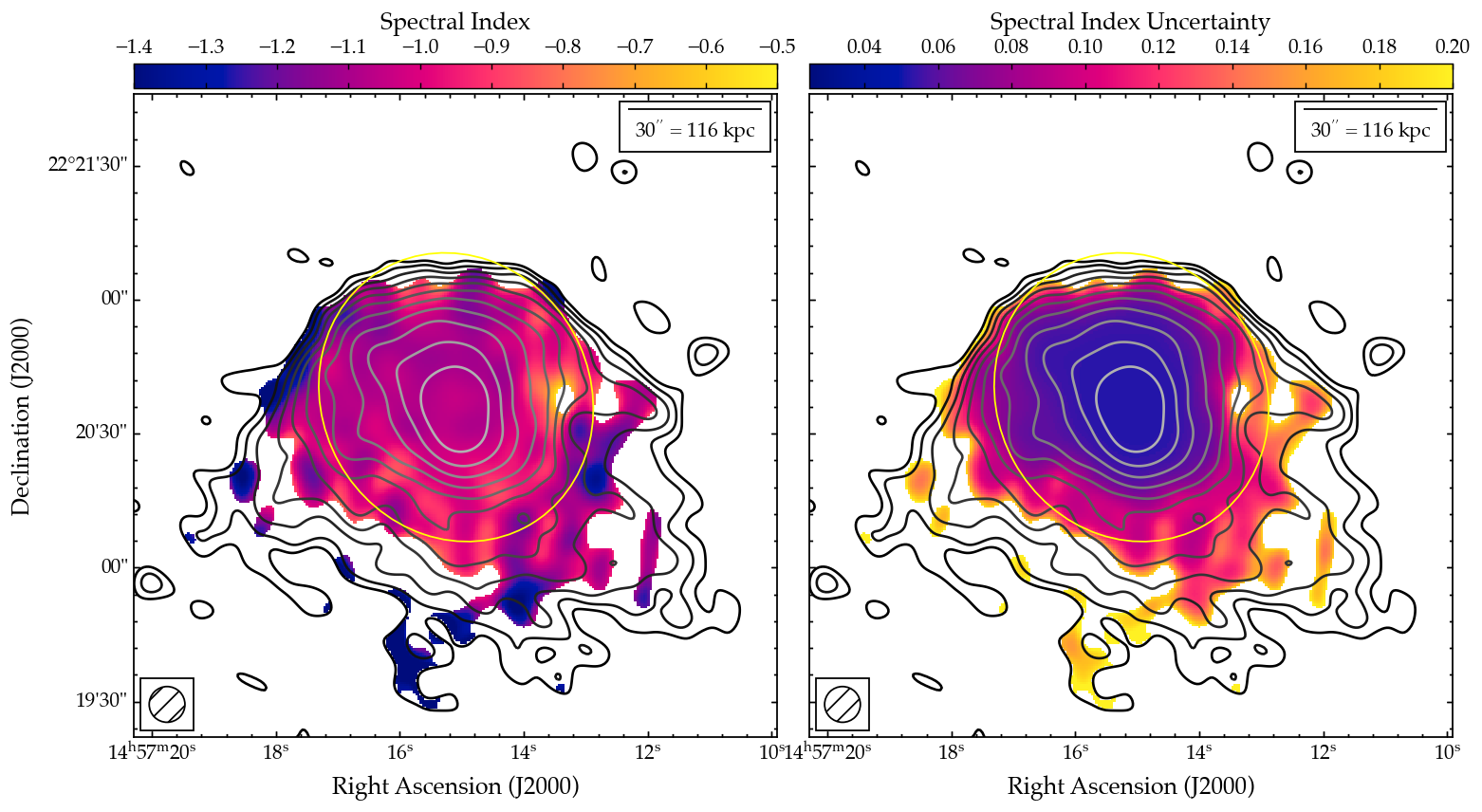}
\includegraphics[width=0.85\linewidth]{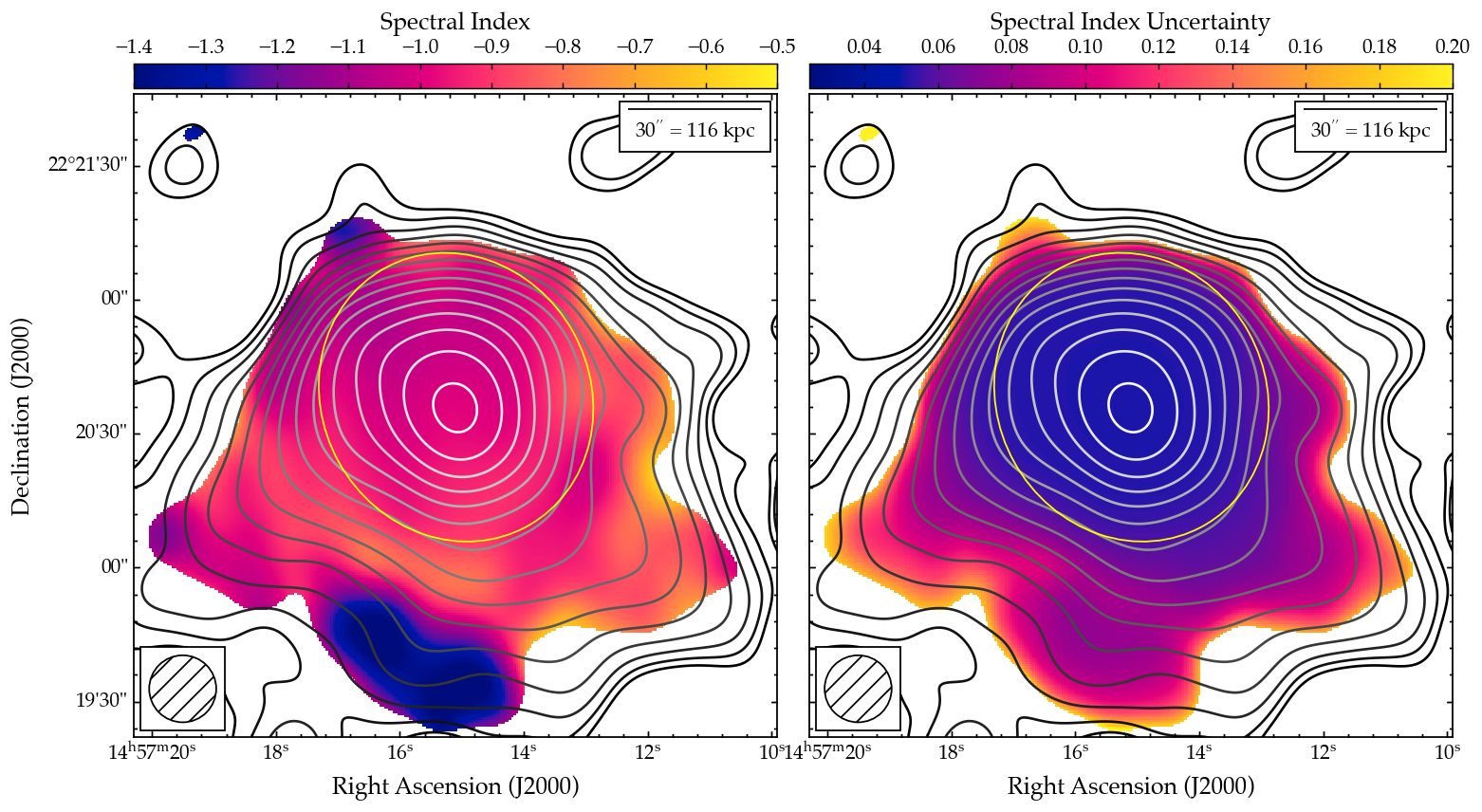}
\cprotect\caption{Spectral index map (\emph{left}) and associated uncertainty (\emph{right}) of the mini-halo in MS~1455.0$+$2232 at a resolution of 8~arcsec (\emph{top}) and 15~arcsec (\emph{bottom}), derived using our MeerKAT (1283~MHz) and LOFAR (145~MHz) data. Contours show MeerKAT surface brightness at 1283~MHz as per Figure~\ref{fig:Z7160_subtracted}. Pixels below $3\sigma$ in each radio image are blanked. The yellow ellipse traces the outer boundary of the sloshing spiral, as per Figure~\ref{fig:radio_xray_overlay}.}
\label{fig:Z7160_subtracted_spectralindex}
\end{center}
\end{figure*}

\subsubsection{Resolved Spectral Properties}\label{sec:alfa_maps}
Figure~\ref{fig:Z7160_subtracted_spectralindex} presents the map of the spectral index between 145~MHz and 1283~MHz at both 8~arcsec and 15~arcsec resolution, as well as the associated uncertainty maps. From the upper panel of Figure~\ref{fig:Z7160_subtracted_spectralindex}, we find a mean spectral index of $\langle \alpha \rangle = -1.07 \pm 0.10$ at 8~arcsec resolution. However, we see reasonably significant fluctuations in the observed spectral index, including some regions where the spectral index appears to steepen, particularly toward the north-eastern and southern/south-western extremes. To investigate the statistical significance of any overall spectral index steepening, we studied the spectral index behaviour both inside and outside the boundaries of the sloshing spiral.

Within the sloshing spiral boundary, we find $\langle \alpha_{\rm in} \rangle = -1.03 \pm 0.08$; for regions outside the sloshing spiral, we find a slightly steeper spectral index of $\langle \alpha_{\rm out} \rangle = -1.11 \pm 0.14$, although these are consistent at the $1\sigma$ level. Similarly, the standard deviation of spectral index is lower interior to the spiral ($\sigma_{\alpha} \sim 0.05$) than outside the spiral boundary ($\sigma_{\alpha} \sim 0.08$). We do note however that this comparison may be limited by small number statistics due to the relative sensitivity of our MeerKAT and LOFAR data.

Our lower-resolution (15~arcsec) spectral index map is presented in the lower panel of Figure~\ref{fig:Z7160_subtracted_spectralindex}. While the mean spectral index of $\langle \alpha \rangle = -0.97 \pm 0.09$ is slightly flatter than the overall mean value we find at 8~arcsec resolution, the two values are consistent to within $1\sigma$. Overall, the spectral index across the majority of the mini-halo is largely uniform when observed at 15~arcsec resolution. 

We repeated the comparison between the observed spectral index interior and exterior to the sloshing spiral at 15~arcsec resolution. As with our higher-resolution map, we find that the average spectral index is slightly steeper exterior to the spiral $(\langle \alpha_{\rm out} \rangle = -1.0 \pm 0.10)$ than interior $(\langle \alpha_{\rm in} \rangle = -0.94 \pm 0.07)$, although again these are consistent at $1\sigma$. The standard deviation in spectral index is also a factor two greater exterior to the spiral ($\sigma_{\alpha} \sim 0.06$) than interior ($\sigma_{\alpha} \sim 0.03$).

Many of the steep-spectrum regions identified at 8~arcsec resolution appear less steep at 15~arcsec resolution --- in particular, the steep-spectrum regions to the north-eastern and south-western edges of the mini-halo flatten slightly from $\alpha \sim -1.4$ to $\alpha \sim -1.1$, which is more comparable with the overall profile. Given the surface brightness sensitivity limit of the available LOFAR data, we cannot currently determine whether the apparent steepening seen at 8~arcsec resolution is the result of signal-to-noise limits, or whether the slight flattening seen at 15~arcsec resolution is the result of blurring different regions with different spectral properties due to the broader PSF FWHM.

Toward the southernmost edge of the mini-halo however, we find a relatively steep spectral index of $\langle \alpha \rangle = -1.45 \pm 0.18$ at 8~arcsec resolution and $\langle \alpha \rangle = -1.37 \pm 0.08$ at 15~arcsec resolution. The fact that the spectral index is consistent in this region at both resolutions suggests that this steepening is physical. Again, we do not currently have the signal-to-noise to investigate this further; deeper low-frequency observations would be required to study the spectral index behaviour in the cluster outskirts in more detail and confirm this steepening.

Due to the sensitivity difference between our MeerKAT and LOFAR images, the outermost regions of the mini-halo are undetected at 145~MHz; however, given that they are undetected by LOFAR they cannot have an ultra-steep spectrum. To quantify the spectral index limit in the outer reaches of the mini-halo, we take the typical surface brightness measured by MeerKAT and assume a $2.5\sigma$ limit from our LOFAR map. This yields an upper limit to the spectral index of $\alpha \gtrsim -1.3$. Deeper observations with LOFAR would be required to map the spectral index in these regions; however, the spectral index in these outer regions is clearly shallower than the ultra-steep values of $\alpha \sim -2$ to $-3$ measured by \cite{Biava2021} in the outer regions of RX~J1720.1$+$2638.

\begin{figure*}
\begin{center}
\includegraphics[width=0.95\linewidth]{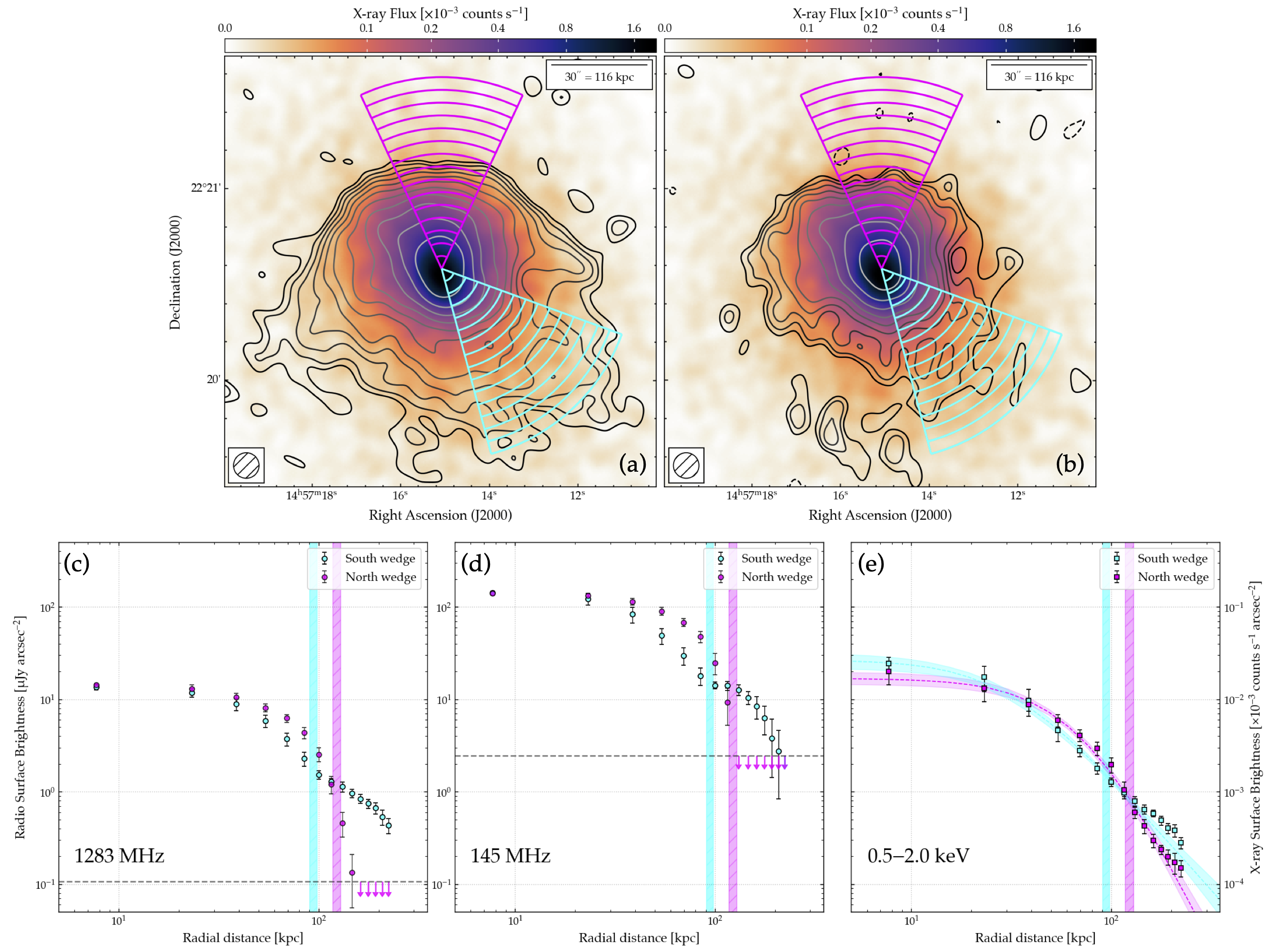}
\cprotect\caption{\emph{Upper:} Radio/X-ray overlays of the mini-halo in MS~1455.0$+$2232. Colourmap in each panel shows our source-subtracted \textit{Chandra} data, smoothed with a Gaussian of FWHM 8~arcsec. Contours show our source-subtracted radio data at 8~arcsec resolution, with MeerKAT 1283~MHz in panel (a) and LOFAR 145~MHz in panel (b). These contours start at $3\sigma$ and scaling by a factor $\sqrt{2}$. The `north wedge' (magenta) and `south wedge' (cyan) extraction regions used to model the radial profiles are also overlaid. \emph{Lower:} azimuthally-averaged radial profiles for the `north wedge' and `south wedge' extracted from our multi-wavelength data, colourised as per panels (a) and (b). Panels (c), (d) and (e) respectively show the profiles extracted from our MeerKAT (1283~MHz), LOFAR (145~MHz) and \textit{Chandra} (0.5$-$2.0~keV) data. Curves in panel (e) denote the best-fit beta model and $1\sigma$ uncertainty. The horizontal dashed line indicates the $1\sigma$ level in the respective radio map. Vertical shaded regions mark the outer edge of the sloshing spiral, as measured along each arc (see Figure~\ref{fig:xray_components}).}
\label{fig:Z7160_radial_profiles}
\end{center}
\end{figure*}

\subsection{Surface Brightness Profiles}
\subsubsection{Radio Surface Brightness}

Given the asymmetric nature of the mini-halo in MS~1455.0$+$2232, we might expect to detect the presence of multiple radio components. Such multi-component nature has only been found in a handful of mini-haloes to-date \citep[][]{Savini2018,Savini2019,Biava2021}. Our observations provide the first hints of radio substructure in this cluster.

We investigated the radial surface brightness profiles of both our radio and X-ray data in our high-resolution (8~arcsec) source-subtracted images. We used two `wedges' centred on the BCG, each covering a 50~deg arc --- a `north wedge' pointing toward the north of MS~1455.0$+$2232 (where the radio contours are tightly packed) and a `south wedge' directed to the south-west where the radio contours are less tightly-packed. Panels (a) and (b) of Figure~\ref{fig:Z7160_radial_profiles} show the wedge regions overlaid atop our MeerKAT 1283~MHz / LOFAR 145~MHz and \textit{Chandra} data.

Panels (c), (d), and (e) of Figure~\ref{fig:Z7160_radial_profiles} show the azimuthally-averaged surface brightness profile for each wedge. At both frequencies, the radio surface brightness profile of each wedge appears to follow a roughly exponential decrease with increasing radius out to about $\sim100$~kpc. However, beyond a distance of around 100~kpc, the behaviour of the radio surface brightness diverges significantly. For the north wedge, the surface brightness decreases sharply; conversely, for the south wedge, a second component emerges. This component also appears to follow a broadly exponentially-decreasing surface brightness profile, and is common to both our MeerKAT and LOFAR data. This provides clear evidence that MS~1455.0$+$2232 hosts a multi-component radio mini-halo.

Typically, a standard circular exponential profile is used to model the surface brightness of haloes and mini-haloes \citep[e.g.][]{Cassano2007,Murgia2009,Bonafede2017}. This profile takes the form:
\begin{equation}\label{eq:efold}
I_{\rm R} (r) = I_{\rm R, c} \exp\left( -r / r_e \right) \, , 
\end{equation} 
where $I_{\rm R, c}$ is the central radio surface brightness and $r_e$ is the $e$-folding radius, typically found to be $\sim R_{\rm H} / 2.6$ for a radio halo of radius $R_\text{H}$ \citep[e.g.][]{Bonafede2017}. However, given the multi-component nature of the diffuse radio emission seen in the south wedge, we can infer that this component likely also underlies the profile seen in the north wedge. In any case, a single exponential does not provide a good description of the radio surface brightness of the mini-halo in MS~1455.0$+$2232.

Recently, \cite{Boxelaar2021} have presented a novel algorithm (the Halo Flux Density CAlculator, or \textsc{Halo-FDCA}) for the measurement of the total flux density from diffuse haloes in galaxy clusters. While this algorithm employs a full MCMC treatment, all models currently implemented by \textsc{Halo-FDCA} rely on the radio surface brightness profile following a single exponential profile\footnote{Whether circular, elliptical, or skewed -- see \url{https://github.com/JortBox/Halo-FDCA}.}. As demonstrated here however, the surface brightness profile of the mini-halo in MS~1455.0$+$2232 cannot be described by a single exponential.

\subsubsection{X-ray Surface Brightness}
The X-ray surface brightness profile recovered by \textit{Chandra} is strikingly different however. It is not immediately obvious whether there is evidence of substructure, as the radial profiles show a similar surface brightness decrease as a function of radius. We attempt to model the behaviour using a standard beta-model, which follows the assumption of an isothermal gas profile and spherical geometry \citep[see e.g.][]{Cavaliere1976,Cavaliere1978,Ettori2000}. This beta model takes the form:
\begin{equation}\label{eq:beta_model}
I_{\rm X}(r) = I_{\rm X, c} \, \left( 1 + \left( \frac{r}{r_{\rm c}} \right)^2 \right)^{\left( 0.5 - 3\beta \right)}, 
\end{equation} 
where $r_{\rm c}$ is the core radius and $I_{\rm X, c}$ is the central X-ray surface brightness. As with the BCG SED, we used \textsc{emcee} to constrain the parameters of Equation~\ref{eq:beta_model}. These fits are also presented in Figure~\ref{fig:Z7160_radial_profiles}.

While the X-ray surface brightness profile of the north wedge is reasonably well-described by a single beta-model, the profile of the south wedge shows an X-ray surface brightness excess beyond around 150~kpc radius. Additionally, we find that the two profiles are not described by the same values of $r_{\rm c}$ or $\beta$: for the north wedge, we find $r_{\rm c, n} = 72.6\pm9.2$~kpc and $\beta_{\rm n} = 0.91 \pm 0.08$, whereas for the south wedge we find $r_{\rm c, s} = 30.9\pm7.2$~kpc and $\beta_{\rm s} = 0.58 \pm 0.06$. This provides further evidence of substructure in the ICM of MS~1455.0$+$2232, and is in agreement with the extension of the diffuse X-ray surface brightness seen in Figure~\ref{fig:xray_components}.

Overall, our evidence indicates that the mini-halo in MS~1455.0$+$2232 shows remarkably different behaviour in azimuthally-averaged radial surface brightness between radio and X-ray wavelengths. To the best of our knowledge, there are no other sufficiently well-resolved mini-haloes where the radial profile has been studied in similar detail. However, we note that this behaviour is in contrast with what has typically been seen for radio haloes, which show a generally similar radial profile at both radio and X-ray wavelengths \citep[e.g.][]{Hoang2021,Rajpurohit2021c}.

\subsection{Thermal/Non-Thermal Comparison}

\subsubsection{Point-to-Point Correlation: Surface Brightness}
In radio haloes, the synchrotron emission often shows a similar morphology and extent to the observed X-ray emission \citep[][and references therein]{vanWeeren2019}. Mini-haloes show a similar trend, albeit covering a smaller fraction of the cluster volume, although that may be due to the relatively limited sensitivity of many historic observations \citep[for example][]{Giacintucci2019,Ignesti2020,Biava2021}. This implies a connection between the thermal and non-thermal components in the ICM. Figure~\ref{fig:radio_xray_overlay} shows the \textit{Chandra} X-ray image of MS~1455.0$+$2232 in the 0.5$-$2.0~keV band, smoothed to a resolution of 15~arcsec, with source-subtracted radio contours overlaid. As can be seen in Figure~\ref{fig:radio_xray_overlay}, the mini-halo in MS~1455.0$+$2232 fills the entire volume of the X-ray emitting region, indicating a strong connection.

We studied the radio/X-ray connection using the point-to-point correlation between the radio surface brightness $(I_{\rm{R}})$ and X-ray surface brightness $(I_{\rm{X}})$ at 15~arcsec resolution. Following recent similar studies \citep[such as][although most of these have focussed on giant haloes]{Ignesti2020,Botteon2020a,Duchesne2021b,Biava2021,Rajpurohit2021b,Rajpurohit2021c}, we place adjacent boxes of 15~arcsec width across the mini-halo in MS~1455.0$+$2232, covering all regions above an X-ray flux limit of $\sim5\times10^{-7}$ counts s$^{-1}$, excising those regions contaminated by residual emission from radio galaxies. These boxes are also shown in Figure~\ref{fig:radio_xray_overlay}. 

\begin{figure*}
\begin{center}
\includegraphics[width=0.95\linewidth]{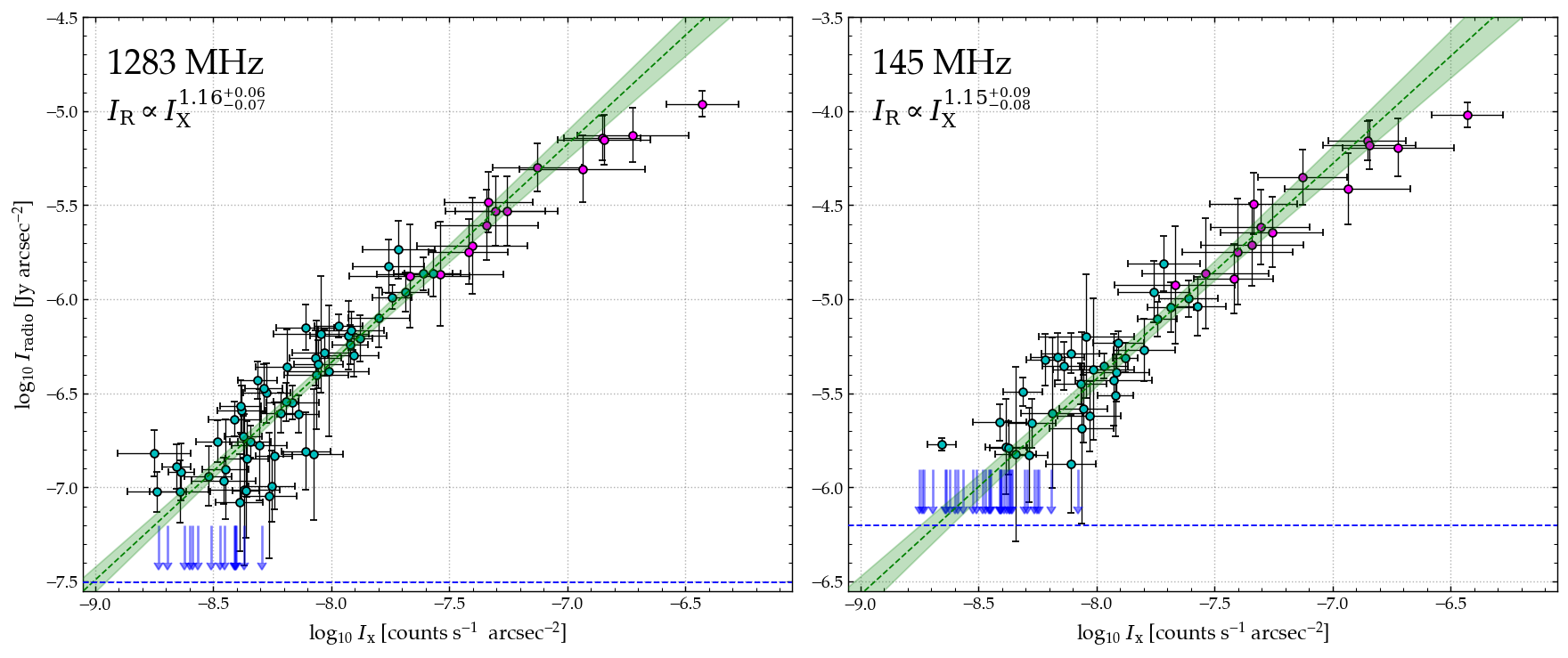}
\cprotect\caption{Radio/X-ray surface brightness correlation for the mini-halo in MS~1455.0$+$2232. \emph{Left panel} shows the $I_{\rm{R}}/I_{\rm{X}}$ plane at 1283~MHz, measured from our MeerKAT image, \emph{right panel} shows the same plane at 145~MHz measured from our LOFAR image, both at 15~arcsec resolution. Datapoint markers are colourised according to the extraction region (seen in Figure~\ref{fig:radio_xray_overlay}). Magenta datapoints lie within the bounds of the sloshing spiral; cyan datapoints lie outside the sloshing spiral boundary. For regions where the radio surface brightness is below the $3\sigma$ level in the corresponding image, we adopt a $2\sigma$ upper limit in this analysis; these are indicated by blue arrows, and the dashed blue indicates the $1\sigma$ level. Dashed green line shows the best-fit power-law relation derived using \textsc{Linmix}, and the shaded green region shows the $1\sigma$ uncertainty. The slope of the best-fit power-law is $b_{\rm{1283~MHz}} = 1.16^{+0.06}_{-0.07}$ at 1283~MHz and $b_{\rm{145~MHz}} = 1.15^{+0.09}_{-0.08}$ at 145~MHz.}
\label{fig:ptp_radio_xray}
\end{center}
\end{figure*}

The $I_{\rm{R}}/I_{\rm{X}}$ plane for the MS~1455.0$+$2232 mini-halo is presented in Figure~\ref{fig:ptp_radio_xray}. Our new highly-sensitive radio data show a positive correlation between radio and X-ray surface brightness, suggesting a strong connection between the thermal and non-thermal components in the mini-halo hosted by MS~1455.0$+$2232. To quantify the connection, we fit a power-law relation (in log-log space) of the form:
\begin{equation}
    {\rm{log}}(I_{\rm{R}}) = c + b {\rm{log}}(I_{\rm{X}}),
\end{equation}
where the slope $b$ describes the scaling between the thermal and non-thermal components of the ICM. The slope of this correlation is related to the underlying particle acceleration mechanism \citep[e.g.][]{Govoni2001,Brunetti2004,ZuHone2013,ZuHone2015}. Broadly-speaking, the turbulent (re-)acceleration scenario can produce either a sub-linear or super-linear slope depending on the nature and distribution of the relativistic electrons throughout the cluster volume. Generally, the hadronic model predicts a super-linear relation due to the central CRp injection profile and equipartition assumption. 

Physically, the slope relates to the relative distribution of non-thermal and thermal components. A super-linear slope (i.e. $b > 1$) would indicate that the magnetic field and/or CRe distribution is more peaked than the thermal plasma distribution, whereas a sub-linear slope (i.e. $b < 1$) would indicate the converse. However, different magnetic field profiles and/or CRp density profiles (which might arise due to the relative contribution from cluster-member galaxies) can yield different correlation slopes \citep[e.g.][]{Ignesti2020,Timmerman2021}.

In the sample of seven mini-haloes studied by \cite{Ignesti2020} with archival GMRT and VLA observations, the point-to-point analysis revealed a consistently super-linear slope in the  $I_{\rm{R}}/I_{\rm{X}}$ plane. However, due to the limited surface-brightness sensitivity and resolution of the available data, the slope of the point-to-point correlations presented by \citeauthor{Ignesti2020} often show significant scatter and thus significant uncertainty.

We determined the best-fit values of $a$ and $b$, and quantified the strength of our correlation, using the \textsc{Linmix} software package\footnote{\textsc{Linmix} is available through \url{https://linmix.readthedocs.io/en/latest/src/linmix.html}} \citep{Kelly2007}. \textsc{Linmix} performs Bayesian linear regression, accounting for uncertainties on both variables, intrinsic scatter and (importantly) upper limits to the dependent variable ($I_{\rm{R}}$). We used the MCMC implementation incorporated in \textsc{Linmix} to determine the best-fit values for $a$ and $b$. We quote the median values of $a$ and $b$ from our MCMC run as our `best-fit' values, and use the 16th and 84th percentiles to determine the uncertainties. Finally, the correlation strength was quantified by measuring the Spearman and Pearson rank coefficients.

Table~\ref{tab:correlation_results} presents the results of our fitting routine. We measure a slope of $b_{\rm{145~MHz}} = 1.15^{+0.09}_{-0.08}$ for the LOFAR/\textit{Chandra} surface brightness plane and $b_{\rm{1283~MHz}} = 1.16^{+0.06}_{-0.07}$ for the MeerKAT/\textit{Chandra} surface brightness plane. We find that the radio and X-ray surface brightness is well-correlated at each frequency, with a Pearson (Spearman) coefficients of $r_{\rm p} = 0.89$ ($r_{\rm s} = 0.95$) at 145~MHz and $r_{\rm p} = 0.91$ ($r_{\rm s} = 0.94$) at 1283~MHz. At each frequency, the slope is super-linear, consistent with the typical behaviour of mini-haloes in the $I_{\rm{R}}/I_{\rm{X}}$ plane \citep[e.g.][]{Govoni2009,Ignesti2020,Biava2021}. This super-linear correlation is generally believed to favour the hadronic model over the turbulent re-acceleration scenario, although we note that under some conditions, the turbulent (re-)acceleration model can replicate a super-linear slope. These conditions include scenarios where turbulence is stronger in the central region and/or where the spatial distribution of CRe is more peaked toward the cluster centre than the thermal electron distribution \citep[see simulations by][for example]{ZuHone2013}.

The mini-halo sample of \cite{Ignesti2020} includes MS~1455.0$+$2232. The analyses are not straightforward to compare due to the dramatic difference in surface brightness sensitivity achieved by our MeerKAT and LOFAR images compared to the narrow-band GMRT images used by \citeauthor{Ignesti2020}. Similarly, the methods employed differ somewhat. \citeauthor{Ignesti2020} use two different methods to study the point-to-point correlation, the results of which exhibit a mild tension: the BCES Bisector method finds a best-fit slope $k = 0.83 \pm 0.10$, whereas they derive a correlation slope of $k = 1.00 \pm 0.12$ from the subsequent Monte-Carlo analysis using the Point-to-point TRend EXtractor \citep[PT-REX;][]{Ignesti2022}. In either case, our \textsc{Linmix} regression yields a fit that is steeper than that found by \cite{Ignesti2020}. 

Our new broad-band data not only allow us to study the point-to-point comparison in exquisite detail, but also to investigate whether there is any evolution in the correlation with frequency. Two of the seven mini-haloes in the sample of \cite{Ignesti2020} have multi-frequency data available at 610~MHz and 1.4~GHz. While both show super-linear slopes at each frequency, one (Abell~3444) shows no change in the correlation slope moving to higher frequency, whereas the other (2A~0335$+$096) shows a slight flattening of the correlation slope with increasing frequency. Recently, LOFAR observations of 2A~0335$+$096 revealed that the correlation slope also shows super-linear scaling at 144 MHz \citep[albeit with large uncertainties, due to the presence of several sub-structures embedded in the mini-halo; for further details see][]{Ignesti2021}.

For radio haloes, few studies have the necessarily high-resolution multi-frequency radio data required to investigate this. However, from these few studies, the picture is also mixed: the radio haloes in Abell~520 and Abell~2744 show no change in the correlation slope with frequency \citep[respectively][]{Hoang2019,Rajpurohit2021c} whereas the radio haloes hosted by ClG~0217+70, MACS~J0717.5$+$3745 and Abell~2256 do exhibit a change in correlation slope with frequency \citep[respectively][Rajpurohit et al. in prep]{Hoang2021,Rajpurohit2021b}. For MACS~J0717.5$+$3745 and Abell~2256, the correlation slope steepens toward higher frequencies, which implies spectral index steepening, in line with the turbulent (re-)acceleration scenario under certain physical conditions. Conversely, ClG~0217+70 exhibits a flatter correlation slope toward higher frequency, suggesting that further non-thermal processes play a role in the cluster outskirts where the X-ray emission is fainter. See the discussion in the aforementioned papers for further details.  

Our analysis indicates no significant evolution in the slope with frequency --- the slope of the correlation is consistent at 1283~MHz and 145~MHz. However, the comparison between our MeerKAT and LOFAR images shows that we cannot yet map the full extent of the mini-halo at 145~MHz; deeper observations with LOFAR, and/or follow-up observations at complementary frequencies would be required to investigate this further. This implies that we are not seeing any evidence of spectral steepening, consistent with evidence presented earlier, although we again note that higher-frequency observations (e.g. at S- or C-band) would be required to investigate further.

\begin{table}
\renewcommand{\arraystretch}{1.4}
\centering
\caption{Summary of results for our \textsc{Linmix} fitting routines, fitting the point-to-point correlation between X-ray surface brightness and either the radio surface brightness at the listed frequency or spectral index, as indicated in the first column. $b$ is the best-fit correlation slope for the plane, and $r_{\rm{S}}$ and $r_{\rm{P}}$ are respectively the Spearman and Pearson correlation coefficient for each plane. The fit for the spectral index point-to-point analysis was performed using only measurements outside the sloshing spiral (cyan datapoints in Figure~\ref{fig:ptp_alfa}). \label{tab:correlation_results}}
\begin{tabular}{lccc}
\hline
Image & Slope & Spearman coeff. & Pearson coeff. \\
      & $b$   & $r_{\rm{S}}$ & $r_{\rm{P}}$ \\
\hline\hline
1283~MHz  & $1.16^{+0.06}_{-0.07}$ &  0.91 & 0.94 \\
145~MHz   & $1.15^{+0.09}_{-0.08}$ &  0.89 & 0.95 \\
\hline
$\alpha$ (outer) & $0.21^{+0.11}_{-0.11}$ &  0.28 & 0.28 \\
\hline
\end{tabular}
\end{table}

\subsubsection{Point-to-Point Correlation: Spectral Index}
Our highly-sensitive data also allow us to study the point-to-point relation between the radio spectral index and the X-ray surface brightness. We also performed this analysis using our images at 15~arcsec resolution. Few such studies exist in the literature for mini-haloes, though \cite{Biava2021} found no strong correlation between spectral index and X-ray surface brightness for the mini-halo in RX~J1720.1$+$2638. Outside of the mini-halo however, \citeauthor{Biava2021} find a moderate-to-strong anticorrelations for the different diffuse radio emission substructures outside the core of RX~J1720.1$+$2638.

For radio haloes, the picture is mixed. Some haloes show positive correlation \citep{Botteon2020b}, some show an anticorrelation \citep{Rajpurohit2021b} and others show both correlation and anticorrelation for different sub-regions \citep{Rajpurohit2021c}. We quantified the relation between radio spectral index $\alpha$ and X-ray surface brightness in by fitting a power-law in linear-log space as follows:
\begin{equation}\label{eq:ptp_alfa}
    \alpha = c + b {\rm{log}}(I_{\rm{X}})
\end{equation}

\begin{figure}
\begin{center}
\includegraphics[width=0.99\linewidth]{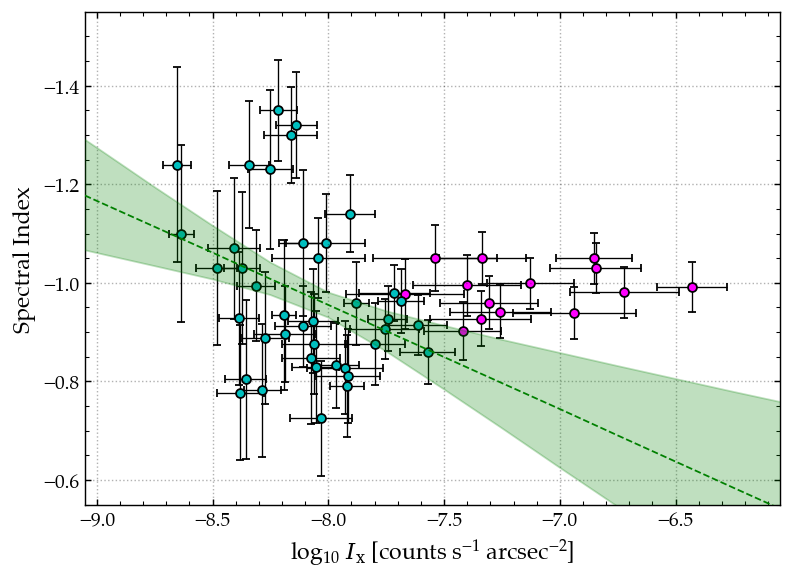}
\cprotect\caption{Radio spectral index/X-ray surface brightness correlation for the mini-halo in MS~1455.0$+$2232. Datapoints are colourised according to extraction region, as per Figure~\ref{fig:radio_xray_overlay}. Magenta points denote regions inside the boundary of the sloshing spiral, cyan points denote regions exterior to the sloshing spiral. Note the inverted $y$-axis, to facilitate comparison with previous similar studies. Dashed green line shows the best-fit to Equation~\ref{eq:ptp_alfa} for the cyan datapoints, derived using \textsc{Linmix}, and the shaded green region shows the $1\sigma$ uncertainty. The slope is $b = 0.21 \pm 0.11$. Magenta points show no correlation.}
\label{fig:ptp_alfa}
\end{center}
\end{figure}

The result of our point-to-point analysis is presented in Figure~\ref{fig:ptp_alfa}, and the correlation slope $b$ and Spearman and Pearson coefficients are reported in Table~\ref{tab:correlation_results}. When we fit a single relation to all datapoints, the slope $b = 0.01^{+0.03}_{-0.03}$ is consistent with no correlation. The Spearman and Pearson coefficients are similarly very weak, as we find $r_{\rm S} = 0.07$ and $r_{\rm P} = -0.01$. However, Figure~\ref{fig:ptp_alfa} suggests that regions inside and outside the sloshing spiral may exhibit different behaviour: regions outside the sloshing spiral (cyan points in Figure~\ref{fig:ptp_alfa}) appear to show a tentative correlation with X-ray surface brightness, similar to the `eastern extension' of the mini-halo in RX~J1720.1$+$2638 \citep{Biava2021}. While the correlation is weak ($r_{\rm S} = 0.28$ and $r_{\rm P} = 0.28$), we find a best-fit slope $b = 0.21^{+0.11}_{-0.11}$. This fit is presented in Figure~\ref{fig:ptp_alfa}. The magenta points show no correlation. 

Figure~\ref{fig:ptp_alfa} also demonstrates that the standard deviation in spectral index increases in the fainter X-ray regions outside the sloshing spiral, as mentioned previously in Section~\ref{sec:alfa_maps}. Overall, this evidence supports the multi-component mini-halo interpretation, though the limited sensitivity of our LOFAR data prevents us from exploring this in more detail at present.

\section{Discussion}\label{sec:discussion}
One of the key unanswered questions in the study of mini-haloes is the underlying acceleration mechanism. In previous sections, we have presented the evidence from our multi-wavelength analysis and examined each piece in isolation. In this section, we will draw all our evidence together and attempt to determine whether the mini-halo in MS~1455.0$+$2232 is consistent with the hadronic scenario or the turbulent (re-)acceleration scenario.

\subsection{Key Observables}
In order to try and differentiate between the underlying acceleration mechanisms, we must examine which scenario `best' reproduces the findings of our analysis. We summarise the key findings below:
\begin{itemize}
    \item The presence of a relatively asymmetric and highly-extended mini-halo $\sim590$~kpc in extent.
    \item An average spectral index of $\alpha \simeq -1$ that is largely consistent over much of the synchrotron-emitting volume. Spectral index fluctuations are seen toward larger radii, although no overall steepening trend is evident (aside from a select region to the south).
    \item The presence of multiple radio components seen in the azimuthally-averaged radial surface brightness profiles to the south of the cluster, but \emph{not} the north.
    \item The existence of a large-scale sloshing spiral (visible in both the X-ray gradient and entropy maps) which does not bound the detected synchrotron emission.
    \item A super-linear scaling relation between X-ray surface brightness and radio surface brightness at both measured frequencies, with a power-law index $b\simeq 1.15$.
\end{itemize}

In both the turbulent (re-)acceleration scenario and the hadronic scenario, the radio emission production density $\dot{\varepsilon}_{\rm R}$ scales approximately with the product of the magnetic field energy densities $(\varepsilon_{B})$ and CRe energy density $(\varepsilon_{\rm CRe})$, respectively, as:
\begin{equation}
\dot{\varepsilon}_{\rm R} \propto \varepsilon_{B} \cdot \varepsilon_{\rm CRe}
\end{equation}
The X-ray emission production density $(\dot{\varepsilon}_{\rm X})$ depends on the electron density $(n_{\rm e})$, gas temperature $(T_{\rm e})$, and thermal gas energy density $(\varepsilon_{\rm th}\propto n_{\rm e} \, T_{\rm e})$ following:
\begin{equation} \label{eq:X-ray-scaling}
\dot{\varepsilon}_{\rm X} ~ \propto ~ n_{\rm e}^{2} \cdot T_{\rm e}^{0.5} ~ \propto ~ \varepsilon_{\rm th}^{2} \cdot T_{\rm e}^{-1.5} 
\end{equation}
Thus, the relation between radio and X-ray emissivities is tied to the dependence of $\varepsilon_{B}$ and $\varepsilon_{\text{CRe}}$ on $n_{\text{e}}$, $\varepsilon_{\text{th}}$, and $T_{\text{e}}$. This relation differs between the two scenarios.

\subsection{Turbulent (Re-)acceleration Scenario}
In this scenario, the CRe are powered by turbulent (re-)acceleration of electrons in the ICM by sloshing motions within the cool core. Our \textit{Chandra} analysis has revealed that MS~1455.0$+$2232 is experiencing large-scale core sloshing on scales up to $\sim1.1$~arcmin ($\sim254$~kpc) in extent; as such, turbulence likely plays a role in powering the mini-halo. The presence of breaks in the radio surface brightness profiles corresponding to the outer boundaries of the sloshing spiral indicates that there must be a strong connection between the dynamics of the thermal gas and the non-thermal components within this region.

Turbulence is inherently an intermittent process, both spatially and temporally. Spatial intermittence implies that the observed radio spectrum along a given line of sight can result from the superposition of different components with different spectral properties. Temporal intermittence implies that the classical steady-state spectra are not necessarily achieved by the CRe distribution at different locations. As such, in the turbulent (re-)acceleration scenario, we might expect to see spatial variations in the radio emission (at a given frequency) and thus fluctuations in the spectral index distribution. 

Our analysis shows that at 15~arcsec resolution, the spectral index is broadly uniform across much of the mini-halo volume. When viewed at 8~arcsec resolution, we see more significant fluctuation. In principle the absence of strong spectral variations may be reconciled with (re-)acceleration assuming that different turbulent cells are integrated along the line of sight; the appearance of spectral variations at higher resolution might be suggestive of this possibility. Furthermore, the average spectral index profile is slightly steeper (and exhibits greater standard deviation) outside the sloshing spiral boundary than within the spiral. This may arise naturally in a scenario where the turbulence is injected by the large-scale sloshing spiral visible in MS~1455.0$+$2232. 

However, the mini-halo extends far beyond the boundaries of the sloshing spiral, and while the X-ray surface brightness shows evidence of a low-level excess following the `tail' of emission to the south/south-west of the mini-halo, we find no significant evidence of any large-scale sloshing in this region. The scaling between radio and X-ray surface brightness is less predictive in this scenario. It depends on the fine details of the turbulence mechanism and how turbulence causes particle acceleration and magnetic field amplification. Simulations of turbulent (re-)acceleration in sloshing cores can yield a super-linear scaling relation \citep{ZuHone2013}, although the extent to which this conclusion depends on the fine-tuning of simulation parameters (dynamical parameters and the turbulent acceleration model) is not explored in the literature.

\subsection{Hadronic Scenario}
In the hadronic scenario, the CRe are produced by inelastic collisions between cosmic ray protons (CRp) and thermal protons in the ICM. Thus, the CRp spectrum imprints onto the observed spectrum of the CRe, as:
\begin{equation}
\dot{\varepsilon}_{\text{CRe}} \propto n_{\rm e} \cdot \varepsilon_{\rm CRp},
\end{equation}
where $\varepsilon_{\rm CRp}$ is the CRp energy density. In galaxy cluster environments, CRp are long-lived and are thus in principle able to achieve a spatially-smooth distribution that exhibits a similar spectral slope throughout the volume of the cluster. As such, the largely-smooth spectral index we observe for this mini-halo is naturally explained by the hadronic scenario. 

The CRe generated by the inelastic collisions are short lived, due to their radiative losses via synchrotron and inverse Compton emission, and therefore should quickly establish a steady state electron energy density according to:
\begin{equation}
\varepsilon_{\rm CRe} ~ = ~ \dot{\varepsilon}_{\rm CRe} \cdot \tau_{\rm loss} ~ \propto ~ \frac{n_{\rm e} \cdot \varepsilon_{\rm CRp}}{\varepsilon_{B} + \varepsilon_{\rm CMB}},
\end{equation}
where $\varepsilon_{\text{CMB}}$ is the CMB energy density and $\tau_{\text{loss}}$ is the timescale over which the electrons radiate their energy. Thus in the hadronic model we expect a relation of the radio emissivity to the basic plasma parameters as:
\begin{eqnarray}
\dot{\varepsilon}_{\rm R} & \propto & \frac{\varepsilon_{B}}{\varepsilon_{B} + \varepsilon_{\rm CMB}} \cdot \varepsilon_{\rm CRp} \cdot n_{\rm th} \nonumber \\
 & \propto & (1+\varepsilon_{\rm CMB} / \varepsilon_{B})^{-1} \cdot \varepsilon_{\rm CRp} \cdot \varepsilon_{\rm th} \cdot T_{\rm e}^{-1}.\label{eq:hadronic-scaling}
\end{eqnarray}
As such, by comparing Equations~\ref{eq:X-ray-scaling} and \ref{eq:hadronic-scaling}, we find that we would achieve a linear scaling between X-ray and radio emission if three conditions are fulfilled:
\begin{itemize}
    \item The energy density of CRp and thermal protons are proportional $(\varepsilon_{\rm CRp} \propto \varepsilon_{\rm th})$.
    \item The core is isothermal ($T_{\rm e}= \rm{const.}$).
    \item The magnetic field energy density is much greater than the energy density of the CMB ($\varepsilon_{B} \gg \varepsilon_{\rm CMB}$, i.e. $B \gg 5~\upmu$G).
\end{itemize}

In case of weaker magnetic fields, with energy densities $\varepsilon_{B} \ll \varepsilon_{\rm CMB}$ roughly scaling with the thermal energy density ($\varepsilon_{B} \propto \varepsilon_{\rm th}$), we would find a strong super-linear scaling relation. In this case, $\dot{\varepsilon}_{\rm R} / \dot{\varepsilon}_{\rm X} \propto \varepsilon_{\rm th} \cdot T_{\rm e}^{0.5}$. As such, assuming a magnetic field in the cluster core of $B \sim B_{\rm IC}$, the observed slightly super-linear scaling relation between radio and X-ray surface brightness arises as a natural consequence of the hadronic model. However, a full investigation of the specific details of the hadronic scenario that are constrained by our observations \citep[similar to, e.g.][]{Ensslin2011,Ignesti2020} is beyond the scope of this paper.

Another important piece of evidence comes from the precipitous drop of the radio brightness observed across the boundary of the sloshing spiral in the north wedge region in Figure~\ref{fig:Z7160_radial_profiles}. In a purely-hadronic scenario, this would require a significant drop of the magnetic field across this boundary, provided that CRp are distributed on large scales in the ICM and that $B \sim B_{\rm IC} \simeq 4-10~\upmu$G. Simulations of the hadronic scenario by \cite{ZuHone2015} do predict a drop in $B$ across the sloshing boundary by a factor 5 or 6; however, in order to explain our observations, a much larger decrease of a factor of at least 30 to 70 in the magnetic field energy density is required (priv.~comm. G.~Brunetti). 

One means by which we can reconcile the hadronic scenario with our observations is to assume that most of the CRp were generated by the central BCG and are confined by the large-scale magnetic field. Thus, the diffusion timescale required is $\tau \sim 5/D_{30}$~Gyr, where $D_{30}$ is the spatial diffusion coefficient in units of $10^{30}$~cm$^{2}$~s$^{-1}$ \citep{Brunetti2014}. As with the turbulent (re-)acceleration scenario, detailed modelling is beyond the scope of this paper.

The evidence provided by our \textit{Chandra} data also cannot be ignored. The large-scale sloshing does not arise in a purely-hadronic scenario. Similarly, we observe both a steepening of the spectral index as well as an increase in standard deviation for regions outside the boundary of the sloshing spiral compared to those regions in the interior. This implies a connection with the large-scale motions within the ICM that is not anticipated in a purely-hadronic scenario.

\subsection{Overall Picture}
Based on the arguments laid out in this section, both the turbulent (re-)acceleration scenario and the hadronic scenario are naturally able to explain many of the observables. However, neither scenario is satisfactorily able to explain all pieces of evidence provided by our detailed multi-wavelength analysis. As such, we are unable to conclusively determine whether one scenario is preferred, or whether a hybrid scenario --- some combination of hadronic emission plus turbulent acceleration, perhaps dominant on different scales \citep[e.g.][]{Brunetti2011,Zandanel2014} --- provides the best explanation. We will return to this in future with the full sample from our MeerKAT-meets-LOFAR mini-halo census, in conjunction with deeper theoretical exploration of the scenarios.

\section{Conclusions}\label{sec:conclusions}
In this paper, we have presented the first highly-sensitive multi-frequency study of the mini-halo hosted by the cluster MS~1455.0$+$2232, using data from the MeerKAT and LOFAR radio telescopes at 1283~MHz and 145~MHz, respectively. These data, combined with archival X-ray data from \textit{Chandra}, have allowed us to perform a comprehensive and detailed study of the connection between the thermal and non-thermal properties of the mini-halo. 

We find that the mini-halo fills the overwhelming majority of the X-ray emitting region traced by \textit{Chandra}, suggesting a strong connection between the thermal and non-thermal properties of the intracluster medium. From our MeerKAT image at 1283~MHz, the largest linear size of this `mini'-halo is 586~kpc, 69 per cent larger than was previously reported, emphasising the need for highly-sensitive radio observations of these relaxed clusters when attempting to understand the thermal/non-thermal connection.

Our re-analysis of deep \textit{Chandra} observations have revealed striking evidence of a large-scale asymmetric sloshing spiral in MS~1455.0$+$2232, identified in the X-ray gradient map. The lower-entropy regions of the ICM follow the structure traced by the sloshing spiral.

We measure a total integrated flux density of $S_{\rm 1283~MHz} = 18.7 \pm 0.9$~mJy and $S_{\rm 145~MHz} = 154.5 \pm 15.6$~mJy for the diffuse emission from the mini-halo. The mini-halo has an integrated spectral index $\alpha^{\rm 145~MHz}_{\rm 1283~MHz} = -0.97 \pm 0.05$. The 2D spectral index map is broadly uniform at 15~arcsec resolution, although it exhibits more significant fluctuations at 8~arcsec resolution, perhaps suggesting the blurring of different substructures by the larger PSF. We find tentative evidence that the average spectral index is marginally steeper outside the boundary of the sloshing spiral; the spectral index exhibits a factor $\sim1.5-2$ greater standard deviation outside the boundary.

We have profiled the behaviour of the radio and X-ray surface brightness, as measured along two arc-shaped wedges from the BCG out to large radius. The radio surface brightness follows a roughly exponential drop-off (although it is poorly-described by standard radio halo profiles) out to the boundaries of the sloshing spiral. Outside the spiral boundary to the north, the radio surface brightness decreases precipitously; to the south, however, we find significant evidence of additional diffuse radio emission indicating the presence of substructure. At present though, we are unable to explore it further.

We have studied the spatial correlation between X-ray surface brightness and (i) radio surface brightness and (ii) radio spectral index. Our analysis reveals a tight correlation in the $I_{\rm{R}}/I_{\rm{X}}$ plane with a conclusively super-linear slope at both 1283~MHz and 145~MHz. We find the slope of correlation (i) to be $b_{\rm 1283~MHz} = 1.16^{+0.06}_{-0.07}$ at 1283~MHz and $b_{\rm 145~MHz} = 1.15^{+0.09}_{-0.08}$ at 145~MHz. For correlation (ii), we find that regions inside the sloshing spiral exhibit no correlation, whereas regions outside the spiral show a relatively weak correlation with a slope $b = 0.21 \pm 0.11$. This provides further evidence in favour of substructure.

Additionally, we have used our new MeerKAT and LOFAR data in conjunction with historic narrow-band radio observations to explore the broad-band radio spectral energy distribution of the brightest cluster galaxy in MS~1455.0$+$2232. The spectral index exhibits a clear break, and is well-described by a broken power-law with a flat low-frequency spectral index $\alpha_{\rm low} = -0.45 \pm 0.05$ and a typical synchrotron spectral index $\alpha_{\rm high} = -0.81 \pm 0.18$ above the break. Due to the relatively sparse sampling of the ancillary data involved however, the break frequency remains relatively poorly constrained, as we find $\nu_{\rm b} = 3.1 \pm 1.4$~GHz.

Finally, we have attempted to determine whether the hadronic model or the turbulent (re-)acceleration scenario is best able to explain all the edivence provided by our observations. While both scenarios are able to explain some observables, neither is satisfactorily able to explain all the evidence. As such, this paper may serve to spark new theoretical work examining the detailed physics of particle acceleration mechanisms in sloshing cool-core clusters. MS~1455.0$+$2232 represents the first of thirteen mini-haloes which our team are studying in detail using deep MeerKAT and LOFAR data; the remainder of our mini-halo census promises to provide a revolutionary step in our understanding of mini-halo physics.

\section*{Acknowledgements}
CJR, NB, A.~Bonafede, EB, and CS acknowledge financial support from the ERC Starting Grant `DRANOEL', number 714245. KR acknowledges financial support from the ERC Starting Grant `MAGCOW', no. 714196. A.~Botteon acknowledges support from the VIDI research programme with project number 639.042.729, which is financed by the Netherlands Organisation for Scientific Research (NWO). FL acknowledges financial support from the Italian Minister for Research and Education (MIUR), project FARE, project code R16PR59747, project name FORNAX-B. FL acknowledges financial support from the Italian Ministry of University and Research $-$ Project Proposal CIR01$\_$00010. RT and RJvW acknowledge support from the ERC Starting Grant ClusterWeb 804208. GDG acknowledges support from the Alexander von Humboldt Foundation. AI acknowledges the Italian PRIN-Miur 2017 (PI A. Cimatti). CS acknowledges support from the MIUR grant FARE `SMS'. CJR thanks Alastair Edge for useful discussions around the high-frequency data available in the literature. We also thank our anonymous referee for their careful and constructive feedback which has strengthened this work.

The MeerKAT telescope is operated by the South African Radio Astronomy Observatory, which is a facility of the National Research Foundation, an agency of the Department of Science and Innovation. CJR wishes to acknowledge the assistance of the MeerKAT science operations team in both preparing for and executing the observations that made this work (and the remainder of project MKT-20126) possible.

LOFAR is the Low Frequency Array designed and constructed by ASTRON. It has observing, data processing, and data storage facilities in several countries, which are owned by various parties (each with their own funding sources), and which are collectively operated by the ILT foundation under a joint scientific policy. The ILT resources have benefited from the following recent major funding sources: CNRS-INSU, Observatoire de Paris and Universit\'e d'Orl\'eans, France; BMBF, MIWF-NRW, MPG, Germany; Science Foundation Ireland (SFI), Department of Business, Enterprise and Innovation (DBEI), Ireland; NWO, The Netherlands; The Science and Technology Facilities Council, UK; Ministry of Science and Higher Education, Poland; The Istituto Nazionale di Astrofisica (INAF), Italy.

This research made use of the Dutch national e-infrastructure with support of the SURF Cooperative (e-infra 180169) and the LOFAR e-infra group. The J\"{u}lich LOFAR Long Term Archive and the German LOFAR network are both coordinated and operated by the J\"{u}lich Supercomputing Centre (JSC), and computing resources on the supercomputer JUWELS at JSC were provided by the Gauss Centre for Supercomputing e.V. (grant CHTB00) through the John von Neumann Institute for Computing (NIC). 

This research made use of the University of Hertfordshire high-performance computing facility and the LOFAR-UK computing facility located at the University of Hertfordshire and supported by STFC [ST/P000096/1], and of the Italian LOFAR IT computing infrastructure supported and operated by INAF, and by the Physics Department of Turin university (under an agreement with Consorzio Interuniversitario per la Fisica Spaziale) at the C3S Supercomputing Centre, Italy.

This research has made use of the VLASS QLimage cutout server at \url{cutouts.cirada.ca}, operated by the Canadian Initiative for Radio Astronomy Data Analysis (CIRADA). CIRADA is funded by a grant from the Canada Foundation for Innovation 2017 Innovation Fund (Project 35999), as well as by the Provinces of Ontario, British Columbia, Alberta, Manitoba and Quebec, in collaboration with the National Research Council of Canada, the US National Radio Astronomy Observatory and Australia’s Commonwealth Scientific and Industrial Research Organisation.

Funding for the Sloan Digital Sky Survey IV has been provided by the Alfred P. Sloan Foundation, the U.S. Department of Energy Office of Science, and the Participating Institutions. SDSS-IV acknowledges support and resources from the Center for High Performance Computing  at the University of Utah. The SDSS website is \url{www.sdss.org}.

SDSS-IV is managed by the Astrophysical Research Consortium for the Participating Institutions of the SDSS Collaboration including the Brazilian Participation Group, the Carnegie Institution for Science, Carnegie Mellon University, Center for Astrophysics | Harvard \& Smithsonian, the Chilean Participation Group, the French Participation Group, Instituto de Astrof\'isica de Canarias, The Johns Hopkins University, Kavli Institute for the 
Physics and Mathematics of the Universe (IPMU) / University of Tokyo, the Korean Participation Group, Lawrence Berkeley National Laboratory, Leibniz Institut f\"ur Astrophysik Potsdam (AIP),  Max-Planck-Institut f\"ur Astronomie (MPIA Heidelberg), Max-Planck-Institut f\"ur Astrophysik (MPA Garching), Max-Planck-Institut f\"ur Extraterrestrische Physik (MPE), National Astronomical Observatories of China, New Mexico State University, New York University, University of Notre Dame, Observat\'ario Nacional / MCTI, The Ohio State University, Pennsylvania State University, Shanghai Astronomical Observatory, United Kingdom Participation Group, Universidad Nacional Aut\'onoma de M\'exico, University of Arizona, University of Colorado Boulder, University of Oxford, University of Portsmouth, University of Utah, University of Virginia, University of Washington, University of Wisconsin, Vanderbilt University, and Yale University.

Finally, we wish to acknowledge the developers of the following python packages (not mentioned explicitly in the text), which were used extensively during this project: \textsc{aplpy} \citep{Robitaille2012}, \textsc{astropy} \citep{Astropy2013}, \textsc{cmasher} \citep{vanderVelden2020}, \textsc{colorcet} \citep{Kovesi2015}, \textsc{matplotlib} \citep{Hunter2007}, \textsc{numpy} \citep{Numpy2011} and \textsc{scipy} \citep{Jones2001}.

\section*{Data Availability}
The images underlying this article will be shared on reasonable request to the corresponding author. Raw MeerKAT visibilities from MKT-20126 are currently proprietary, with public release nominally scheduled for December 2022; upon release, they can be accessed via the SARAO archive (\url{https://apps.sarao.ac.za/katpaws/archive-search}). Raw LOFAR visibilities can be accessed via the LOFAR Long-Term Archive (LTA; \url{https://lta.lofar.eu}). The \textit{Chandra} data are available via the Chandra Data Archive (\url{https://cxc.harvard.edu/cda/}).

\bibliographystyle{mnras}
\bibliography{MS1455_MeerKAT-meets-LOFAR}

\appendix

\section{Value-Added Data For Subtracted Sources}

\begin{table*}
\footnotesize
\centering
\caption{Flux density measurements at 1283~MHz and 145~MHz, as well as the corresponding radio spectral index $\alpha$ for sources identified with yellow `+' signs in Figure~\ref{fig:Z7160_continuum}. Cross-identifications with SDSS optical counterparts are listed where available, along with redshift measurements ($z$), both taken from SDSS DR16 \citep{Ahumada2020}. We quote spectroscopic redshifts where present, photometric redshifts otherwise (identified with a superscript `$p$'). Galaxies where the association and/or the SDSS photometry is uncertain are marked by a superscript `$u$'. \label{tab:measurement}}
\begin{tabular}{lrrcccrl}
\hline
ID & Right Ascension & Declination & $S_{\rm 1283~MHz}$ & $S_{\rm 145~MHz}$ & $\alpha$ & SDSS ID & $z$ \\
   & (J2000) & (J2000) & [mJy] & [mJy] & & &  \\
\hline\hline 
1  & 14:57:08.16 & +22:21:09.0 & $0.580 \pm 0.153$ & $2.30 \pm 0.25$ & $-0.63 \pm 0.13$ & SDSS~J145708.15+222109.4 & 0.124 \\
2  & 14:57:11.36 & +22:20:55.6 & $0.065 \pm 0.007$ & $-$ & $-$ & $-$ & $-$ \\
3  & 14:57:14.53 & +22:19:41.4 & $0.043 \pm 0.004$ & $-$ & $-$ & SDSS~J145714.42+221943.0 & 0.290 \\
4  & 14:57:14.73 & +22:19:32.6 & $0.077 \pm 0.010$ & $0.46 \pm 0.06$ & $-0.82 \pm 0.08$ & SDSS~J145714.74+221933.2 & 0.513$^{p}$ \\
5  & 14:57:15.13 & +22:18:54.2 & $0.113 \pm 0.006$ & $-$ & $-$ & SDSS~J145715.13+221854.3 & 0.466$^{p}$ \\
6  & 14:57:15.20 & +22:19:15.0 & $0.048 \pm 0.003$ & $-$ & $-$ & $-$ & $-$ \\
7  & 14:57:16.10 & +22:19:49.2 & $0.212 \pm 0.029$ & $1.05 \pm 0.14$ & $-0.73 \pm 0.09$ & SDSS~J145716.09+221950.6 & 0.264 \\
8  & 14:57:16.15 & +22:19:04.9 & $0.053 \pm 0.005$ & $-$ & $-$ & SDSS~J145716.12+221905.8 & 0.396$^{p}$ \\
9  & 14:57:16.31 & +22:19:40.4 & $0.144 \pm 0.012$ & $0.88 \pm 0.13$ & $-0.83 \pm 0.08$ & $-$ & $-$ \\
10 & 14:57:16.64 & +22:21:16.3 & $0.054 \pm 0.003$ & $-$ & $-$ & $-$ & $-$ \\
11 & 14:57:16.72 & +22:20:55.6 & $0.166 \pm 0.012$ & $-$ & $-$ & SDSS~J145716.72+222055.4 & 0.433$^{p,u}$ \\
12 & 14:57:17.72 & +22:19:22.2 & $0.325 \pm 0.002$ & $2.22 \pm 0.25$ & $-0.88 \pm 0.05$ & $-$ & $-$ \\
13 & 14:57:18.21 & +22:19:06.3 & $0.151 \pm 0.009$ & $1.05 \pm 0.02$ & $-0.88 \pm 0.03$ & SDSS~J145718.59+221906.6 & $-$ \\
14 & 14:57:18.39 & +22:21:40.1 & $0.024 \pm 0.002$ & $-$ & $-$ & $-$ & $-$ \\
15 & 14:57:18.48 & +22:18:57.2 & $0.031 \pm 0.004$ & $-$ & $-$  & $-$ & $-$ \\
16 & 14:57:19.00 & +22:19:56.4 & $0.042 \pm 0.004$ & $-$ & $-$ & $-$ & $-$ \\
17 & 14:57:19.13 & +22:21:20.7 & $0.047 \pm 0.003$ & $-$ & $-$ & $-$ & $-$ \\
18 & 14:57:19.38 & +22:21:34.9 & $0.111 \pm 0.006$ & $1.09 \pm 0.12$ & $-1.04 \pm 0.05$ & SDSS~J145719.37+222135.6 & 0.770$^{p}$ \\
\hline
\end{tabular}
\end{table*}

\bsp	
\label{lastpage}
\end{document}